\documentclass[a4paper, 12pt]{article}
\usepackage{graphicx}
\usepackage{amsmath}
\usepackage{amssymb}
\usepackage{subfigure}

\def \vs  {\vskip5mm}
\def \be {\begin{equation}}
\def \ee {\end{equation}}
\def \bea {\begin{eqnarray}}  
\def \eea {\end{eqnarray}} 
\def \mea {\nonumber\\}
\def \half {{\textstyle \frac{1}{2}}}
\def \quarter {{\textstyle \frac{1}{4}}}

\begin{document}    
\begin{titlepage}

\title{B\"acklund flux-quantization in a  model of electrodiffusion based on Painlev\'e II}     
     
\author{
A.J. Bracken$^1$,
\footnote{{\em Email:} a.bracken@uq.edu.au}\,
L. Bass$^1$
\footnote{{\em Email:} lb@maths.uq.edu.au} 
and C. Rogers$^{2,\, 3}$
\footnote{{\em Email:} c.rogers@unsw.edu.au}
\\$^1$ Department of Mathematics\\    
The University of Queensland\\Brisbane, Australia\\$^2$ Department of Applied Mathematics\\The Hong Kong Polytechnic University\\Hong Kong\\$^3$ Australian Research Council Centre of Excellence \\for Mathematics \& Statistics of Complex Systems\\School of Mathematics and Statistics\\The University of New South Wales\\Sydney, Australia}

\date{}     
\maketitle     
\vs
\begin{abstract}
A previously-established
model of steady 
one-dimensional two-ion electrodiffusion across a liquid junction is reconsidered.  It involves three coupled first-order nonlinear 
ordinary differential equations, and has the second-order Painlev\'e II equation at its core.  
Solutions are now grouped by B\"acklund transformations into infinite sequences, partially labelled by two 
B\"acklund invariants. 
Each sequence is characterized by evenly-spaced quantized fluxes of the two
ionic species, and hence evenly-spaced quantization of the electric current-density. Finite subsequences of exact solutions
are identified,  
with positive ionic concentrations and quantized fluxes, starting from a solution with zero electric field found by Planck, 
and suggesting an interpretation as a ground state plus excited states of the system.   
Positivity of ionic concentrations 
is established  whenever Planck's charge-neutral boundary-conditions apply. 
Exact solutions are obtained for the electric field and ionic concentrations in well-stirred reservoirs 
outside each face of the junction, enabling the formulation of more realistic boundary-conditions. 
In an approximate form, 
these lead to radiation boundary conditions for Painlev\'e II. Illustrative numerical solutions are presented, and
the problem of establishing compatibility of boundary conditions with the structure of 
flux-quantizing sequences is discussed.   
\end{abstract}

     
\end{titlepage}

\section{Introduction}
Transport of charged ions across liquid junctions plays a fundamental role in a variety of important 
physical and biological contexts, in particular semiconductor theory, electrochemistry and models of the
nervous system.  Many years ago, an extension of the Nernst-Planck model of ion transport \cite{nernst,planck} 
was proposed \cite{grafov,bass1,bass2}, incorporating the effect of the
electric field that develops within the junction in response to charge separation.  Such separation tends to develop 
as ions diffuse with different diffusion coefficients, while the consequent electric 
field acts  to inhibit it.  This feedback  leads to nonlinearity of the extended model. 

Mathematically, the extension involves the 
use of Gauss' Law to relate charge density to electric field, together with 
the imposition of
Einstein's relation between diffusivities and mobilities of the ions.  In its simplest form, the model
deals with one-dimensional transport across an infinite slab occupying $0\leq x \leq \delta$, 
for two types of ions carrying equal and opposite charges. 
With the concentrations of the two ionic species 
and 
the induced electric field within the slab denoted by $c_{\pm}(x)$ and $E(x)$ respectively, 
the governing system of coupled ordinary differential equations (ODEs) 
obtained is
\bea
c_+\,'(x)= ({\tilde z}e/k_B T)\,E(x)\,c_+(x)-\Phi_+/D_+\,,
\mea\mea
c_-\,'(x)= -({\tilde z}e/k_B T)\,E(x)\,c_-(x)-\Phi_-/D_-\,,
\mea\mea
E\,'(x)=(4\pi {\tilde z} e/\epsilon)\left[c_+(x)-c_-(x)\right]\qquad
\label{system1}
\eea
for $0<x<\delta$.  Here $\Phi_+$ and $\Phi_-$ denote the steady 
(constant) fluxes of the two species in the $x$-direction across the slab, ${\tilde z}$ their common valence, 
and $D_{\pm}$ the corresponding diffusion constants, while $k_B$ denotes Boltzmann's constant, $e$
the electronic charge, and $T$
the ambient absolute temperature within the solution in the slab.  
An important auxiliary quantity is the 
electric current-density
\bea
J={\tilde z}e\,(\Phi_+-\Phi_-)
\,.
\label{current1}
\eea

The slab equations \eqref{system1} when supplemented by two-point boundary conditions (BCs) specify the 
electrical structure of liquid junctions including nerve membranes.  The latter are so thin (of the order of
$10^{-6}\,cm$) that typical physiological potential differences of $0.1\, V$ generate large fields of order $10^5\,V/cm$. 
In that context it is not only the potential difference across the slab (which was of primary interest to Nernst \cite{nernst} and Planck \cite{planck}) but the electric field distribution within it that may be important \cite{bass3}.  

In a special case when $\Phi_-=0$, it was shown \cite{grafov} that as a consequence of \eqref{system1}, $c_-(x)$ satisfies a second order
nonlinear equation 
that can be transformed into the Painlev\'e II ODE (PII).  Independently, it was shown \cite{bass1} that in the general case, and with 2-point BCs, 
the field $E(x)$ also satisfies a form of PII (see \eqref{painleve1} below).  These were perhaps the first applications of PII to 
physics or biology.  Since the introduction of the model defined by \eqref{system1}, this aspect and others have been 
much studied (see \cite{rubinstein,ben,zaltzman} and references therein).  In particular, more recent analysis \cite{{rogers1,rogers6}} 
has drawn attention to the relevance to the model of 
already-known exact solutions of PII, and of B\"acklund transformations \cite{rogers7} of its solutions more generally.     

In the formulation given in \cite{bass1}, the model equations (\ref{system1}) were supplemented by BCs 
corresponding to
charge-neutrality at each face of the slab, so that $c_+(0)=c_-(0)$ and $c_+(\delta)=c_-(\delta)$.  From \eqref{system1}, these imply that 
$E\,'(0)=0=E\,'(\delta)$, and hence lead to BCs of Neumann type for PII as it appears in \eqref{painleve1} below).    
We shall discuss the relevance of these BCs, and the possibility of imposing others corresponding to the presence of 
well-stirred ionic reservoir solutions
outside the membrane slab, occupying $-\infty<x<0$ and $\delta<x<\infty$.  We shall also downplay the role of PII, 
emphasized in previous 
analyses of the model, 
in part because of the peculiar feature that 
boundary values of $E(x)$ appear in the coefficients of that  ODE  for $E(x)$ itself, as it appears
in the form \eqref{painleve1}.
This complicates its theoretical and numerical analysis \cite{thompson,amster}. In contrast, we find that the system \eqref{system1} is 
solved numerically with no difficulty 
by commercial packages such as MAPLE \cite{maple}, at least in cases corresponding to charge-neutrality at 
$x=0$ and $x=\delta$. (See Sec. 6.1.)   

By working
directly with the system \eqref{system1}, we also avoid singling out the electric field, and give comparable weight to 
the ionic concentrations, and their associated fluxes.  This will enable us to elucidate 
the role of 
B\"acklund transformations in producing sequences of solutions of \eqref{system1} in which the fluxes are quantized.    

Of particular interest are finite subsequences of exact solutions with positive concentrations, generated by B\"acklund transformations from  a 
solution of \eqref{system1}, first written down by Planck \cite{planck},  with $c_+(x)=c_-(x)$ and $E(x)=0$ throughout the junction.
It is remarkable that Planck's solution, obtained by him when modelling electrodiffusion 10 years before his fundamental work on 
quantization of black-body radiation, should now appear some $120$ years later as the seed solution 
in sequences exhibiting B\"acklund flux-quantization. 

As we shall see, there are difficulties reconciling either of the two kinds of BCs mentioned above, 
with the formation of such sequences of solutions.  This raises the problem of identifying 
what form of BCs if any is consistent with  `B\"acklund flux-quantization.'
\section{Preliminary remarks and Painlev\'e II} 
With the introduction of a suitable constant reference concentration $c_{ref.}$, to be identified later, 
and with the change to dimensionless independent and dependent variables
\bea
x^*=x/\delta\,,\quad c_{\pm}^*(x^*)=c_{\pm}(x)/c_{ref.}\,,
\mea\mea
E^*(x^*)=({\tilde z}e\delta/k_BT)\,E(x)\,,\qquad
\label{change_variables}
\eea
and dimensionless constants
\bea
A_{\pm}=-\Phi_{\pm}\delta /c_{ref.} D_{\pm}\,,\quad \lambda=\sqrt{\epsilon k_B T/4\pi({\tilde z}e)^2\delta ^2\,c_{ref.}}
\label{change_constants}
\eea  
the equations \eqref{system1} become (on immediately dropping all the asterisks)
\bea
c_+\,'(x)= E(x)\,c_+(x)+A_+\,,
\mea\mea
c_-\,'(x)= -E(x)\,c_-(x)+A_-\,,
\mea\mea
\lambda^2 E\,'(x)=c_+(x)-c_-(x)
\label{system2}
\eea 
for $0< x<1$. The current density is also conveniently made dimensionless by setting
\bea
j=\frac{\delta}{({\tilde z}e)c_{ref.}(D_++D_-)}\,J\,.
\label{current2}
\eea
In terms entirely of dimensionless quantities we then have 
\bea
j= \alpha_-A_--\alpha_+ A_+\,,
\label{current3}
\eea
where
\bea
\alpha_{\pm}=D_{\pm}/(D_++D_-)\,\,\,({\rm so\,\,that}\,\,\alpha_{\pm}>0\,\,{\rm and}\,\,\alpha_++\alpha_-=1)\,.
\label{alpha_defs}
\eea   

After the first two of \eqref{system2} are added, and the third is applied, 
one integration can be performed to deduce that
\bea
P(x)-\theta x= B\quad{\rm (const.)}\,,
\label{first_integral}
\eea
where
\bea
P(x)= c_+(x)+c_-(x)- \half \lambda ^2 E(x)^2\,,\quad \theta=A_++A_-\,.
\label{Pdef}
\eea

Eq. \eqref{first_integral} has the general meaning of a work-energy relation.  
This can be seen by returning to the original variables for the moment and multiplying through by $k_BT$, to obtain the result
\bea
k_B T[c_+(x)+c_-(x)]-\epsilon E(x)^2/8\pi-[\Phi_+/u_+ +\Phi_-/u_-]x= {\rm const.}
\label{first_integral2}
\eea
where $u_+=D_+/k_B T$ and $u_-=D_-/k_B T$ are the mobilities of the two types of ions.  
Here $k_B T[c_+(x)+c_-(x)]-\epsilon E(x)^2/8\pi$ is the ionic (osmotic) pressure at $x$ in the presence of the electric field.  The minus sign
reflects the fact that the electric field tends to keep the oppositely charged ions together, effectively reducing their degrees of freedom
and hence the associated pressure.   
The term $[\Phi_+/u_+ ]x$ (resp. $[\Phi_-/u_-]x$) is the work done per unit volume 
against the resistance offered by the background fluid, in transporting positive (resp. negative) ions from $0$ to $x$.
Accordingly, we may refer to $P(x)$ and $\theta$ as (the dimensionless forms of) the pressure at $x$, and the resistive force-density, 
respectively.        
 
Differentiating the third of \eqref{system2} once and then applying the first two of \eqref{system2} to eliminate
$c_+\,'(x)$ and $c_-\,'(x)$ from the result, leads to 
\bea
\lambda ^2 E\,''(x)=E(x)\,[c_+(x)+c_-(x)]+(A_+-A_-)
\label{interim}
\eea
and, with the help of \eqref{first_integral} to 
\bea
\lambda ^2E\,''(x)-\half \lambda ^2 E(x)^3-(\theta x+B) E(x)-(A_+-A_-)=0\,.
\label{painleve1}
\eea
This can be brought to a standard form of PII by a constant scaling of the dependent variable, 
together with
a linear transformation of the independent variable
\cite{bass1,rogers6}. 

From any solution $E(x)$ of \eqref{painleve1}, corresponding 
$c_{\pm}(x)$ can be obtained using \eqref{first_integral} and the third of 
\eqref{system2}.  In this way, PII determines all solutions of 
\eqref{system2}.  The results of many studies \cite{rogers1}-\cite{amster}, \cite{painleve}-\cite{tsuda} of PII
can then be brought to bear on the problem of interest, in particular 
B\"acklund transformations and exact solutions.  These have been extensively explored  elsewhere \cite{rogers6} 
in the context of electrodiffusion on a half line rather than a finite interval of the $x$-axis,
which is our focus here.
\section{Planck's solution}
Planck \cite{planck} considered the situation where $\lambda$ is vanishingly small.  As \eqref{change_constants} shows, this occurs when the electric charges 
$\pm {\tilde z}e$ are much larger in magnitude than $\sqrt{\epsilon k_BT/4\pi\delta^2\,c_{ref.}}$.
(Here the choice of $c_{ref.}$ becomes relevant --- see the {\em Remark} following \eqref{reservoir9} below.)    
In the limit $\lambda \to 0$, with $c(0)=c_0$, the system \eqref{system2} is easily solved
to give Planck's solution
\bea
c_+(x)=c_-(x)=c(x)=c_0+\half(A_++A_-)x\,,
\mea\mea
E(x)=\half (A_--A_+)/c(x)\,.\qquad\qquad
\label{planck}
\eea
With the help of \eqref{current1} the electric field may also be written as \cite{bass2}
\bea
E(x)=\frac{j}{c(x)}+\frac{(\alpha_+-\alpha_-)(c_1-c_0)}{c(x)}\,,
\label{planck2}
\eea
where $c_1=c(1)=c_0+(A_++A_-)/2$, 
which brings to the fore the role of the electric current density: the first term on the RHS of \eqref{planck2} is the ohmic
contribution to the field, and the second term is the gradient of the Nernst-Planck potential \cite{planck}.  The situation where $j=0$ is of particular physical interest.

For any value of $j$, Planck's solution \eqref{planck}  might be used as the zeroth-order term 
in an asymptotic expansion, in powers of $\lambda ^2$, 
of a solution to
the system \eqref{system2} with $\lambda \neq 0$.  This would require the methods of singular perturbation theory,
as is clear from the way in which $\lambda ^2$ appears multiplying the highest derivative in  \eqref{painleve1}.   
That approach has  been developed previously for closely related problems \cite{macgillivray,jackson}, 
and will not be pursued here.     

For our purposes in what follows, it is more important to note that \eqref{planck} defines an exact solution of the full system \eqref{system2}
in the special 
case that  $A_+=A_-=c_1-c_0$ and hence $E(x)\equiv 0$.  This solution, which has $j\neq 0$, plays the role of the seed for 
an infinite sequence of exact rational solutions \cite{yablonskii,vorobev,clarkson,noumi,rogers1,rogers6} generated by B\"acklund transformations,
as highlighted in Sec. 6 below.  
\section{A discrete symmetry group}
The system \eqref{system2}  must be supplemented by appropriate BCs, as we discuss below, to determine 
physically meaningful solutions with nonsingular $c_{\pm}(x)$, positive on $[0,1]$.  For the time being we ignore
such considerations.  Bearing in mind the way that PII determines all solutions of \eqref{system2}, and that solutions of PII itself, when
regarded as analytic functions, may have (movable) simple poles, we mean by  
``a solution ${\cal S}=(c_+\,,c_-\,,E\,,A_+\,,A_-)$",
any two real constants $A_+$, $A_-$, together with any three real-valued 
functions $c_+$, $c_-$, \,$E$ 
satisfying (\ref{system2}) almost everywhere on $[0,1]$.

Given one solution ${\cal S}=(c_+\,,c_-\,,E\,,A_+\,,A_-)$, straightforward if sometimes lengthy 
manipulations 
confirm that 
in general there exist four other solutions.  These transformed solutions can be found 
from known symmetries of PII \cite{rogers6}, 
but they are presented together here, in terms of the variables of the system \eqref{system2},
for the first time.  The transformed solutions, each of which we denote by 
${\hat {\cal S}}=({\hat c}_+\,,{\hat c}_-\,,{\hat E}\,,{\hat A}_+\,,{\hat A}_-)$ 
in turn, are as follows :---   
\subsection{Conjugate solution}

Here ${\hat {\cal S}}={\cal C}({\cal S})$, where
\bea 
{\hat c}_+=c_-\,,\quad {\hat c}_-=c_+\,,\quad {\hat E}=-E\,,
\quad
{\hat A}_+=A_-\,,\quad {\hat A}_-=A_+\,.
\label{conjugate}
\eea

Noting that a second application of ${\cal C}$ returns ${\hat {\cal S}}$ to the starting solution ${\cal S}$, we write
\bea
{\cal C}^2=I\,.
\label{conjugate_squared}
\eea
\subsection{Reflected solution}
Here ${\hat {\cal S}}={\cal R}({\cal S})$, where
\bea 
{\hat c}_+(x)=c_+(1-x)\,,\quad {\hat c}_-(x)=c_-(1-x)\,, 
\mea\mea
{\hat E}(x)=-E(1-x)\,,\quad
{\hat A}_+=-A_+\,,\quad {\hat A}_-=-A_-\,.
\label{reflection}
\eea
We note that  
\bea
{\cal R}^2=I\,,\quad {\cal R}{\cal C}={\cal C}{\cal R}\,.
\label{CP_relations}
\eea
\subsection{B\"acklund-transformed solution}
Here ${\hat {\cal S}}={\cal B}({\cal S})$, where
\bea 
{\hat c}_+=c_-+2\lambda^2 A_+ E/ c_+ +2\lambda^2 A_+^2/ c_+^2\,,\quad {\hat c}_-=c_+\,,
\mea\mea
{\hat E}=-E-2A_+/c_+\,,\quad {\hat A}_+=2A_++A_-\,,\quad {\hat A}_-=-A_+\,.
\label{backlund}
\eea
We note that if $A_+=0$, then ${\cal B}(S)={\cal C}(S)$.
\subsection{Inverse B\"acklund-transformed solution}
Here ${\hat {\cal S}}={\cal B}^{-1}({\cal S})$, where
\bea 
{\hat c}_+=c_-\,,\quad {\hat c}_-=c_+-2\lambda^2 A_- E/ c_- +2\lambda^2 A_-^2/ c_-^2\,,
\mea\mea
{\hat E}=-E+2A_-/c_-\,,\quad {\hat A}_+=-A_-\,,\quad {\hat A}_-=2A_-+A_+\,.
\label{backlund_inverse}
\eea
We note that if $A_-=0$, then ${\cal B}^{-1}({\cal S})={\cal C}({\cal S})$.

It is straightforward to check that ${\cal B}^{-1}$ is indeed inverse to ${\cal B}$, and we write
\bea
{\cal B}^{-1}{\cal B}=I={\cal B}{\cal B}^{-1}\,.
\label{inverse_relation}
\eea

It is also easily checked that
\bea
{\cal C}{\cal B}={\cal B}^{-1}{\cal C}\,,\quad {\cal R}{\cal B}= {\cal B}{\cal R}\,,\quad {\cal R}{\cal B}^{-1}
= {\cal B}^{-1}{\cal R}\,.
\label{CRB_relations}
\eea
From \eqref{CRB_relations}, \eqref{CP_relations} and \eqref{conjugate_squared}, it then follows that ${\cal C}$,
${\cal R}$ and ${\cal B}$, together with their inverses and with $I$ as unit, 
generate a discrete invariance group of transformations
of the system \eqref{system2}. Except on special solutions, this group is of infinite order.  

From 
\eqref{conjugate}, \eqref{backlund} and \eqref{backlund_inverse}, it is also easily seen that the quantities $P(x)$ 
(pressure at $x$) and $\theta$ 
(resistive force density) as in \eqref{first_integral} and \eqref{Pdef} are invariant under the action of the subgroup generated by 
${\cal C}$, ${\cal B}$ and ${\cal B}^{-1}$.  It is unclear why these physical quantities should be singled out to be invariants. 
The invariant $\theta$ plays a central role in flux-quantization below.    
\subsection{Gambier-transformed solutions}
There also exist invertible  transformations, which we  call Gambier and inverse Gambier transformations,   that 
carry  solutions back and forth between two special subclasses.  These
subclasses are of particular interest because one 
contains, among others, exact solutions of rational type, while the other contains, among others, exact solutions of Airy type \cite{rogers6}. 
These
transformations and their inverses, to be denoted by ${\cal G}_{\pm}$ and ${\cal G}_{\pm}^{(-1)}$ respectively, 
and to be defined in what follows, 
can also be derived from a known (restricted) symmetry of PII 
discovered by Gambier
\cite{gambier,witte,tsuda}. 
We present them explicitly here 
in terms of the variables in \eqref{system2} for the first time, as follows.

The transformation ${\cal G}_+$ (resp. ${\cal G}_-$) converts any solution ${\cal S}$ in which  $A_+=0$
(resp. $A_-=0$), 
into a solution ${\hat {\cal S}}$ in which ${\hat A_+}={\hat A_-}$.  If ${\cal S}$ has the value $B$ for the  
first-integral  in \eqref{first_integral}, then ${\hat{\cal S}}$ has corresponding 
value 
${\hat B}=-B/2$.  
In detail, if $A_+=0$ and $A_-=A$, the transformed solution ${\hat {\cal S}}={\cal G}_+({\cal S})$ has  
\bea
{\hat c}_+(x)=\quarter \lambda E(x)\sqrt{2c_+(x)}+\half c_+(x)-\quarter (Ax+B)\,,
\mea\mea
{\hat c}_-(x)=-\quarter\lambda  E(x)\sqrt{2c_+(x)}+\half c_+(x)-\quarter (Ax+B)\,,
\mea\mea
{\hat E}(x)=\sqrt{2c_+(x)}\,/\lambda\,,\quad {\hat A}_+={\hat A}_-=-\quarter A\,,\quad {\hat B}=-\half B\,.
\label{gambier1}
\eea
Similarly, if $A_-=0$ and $A_+=A$, the transformed solution ${\cal G}_-({\cal S})$ has  
\bea
{\hat c}_+(x)=-\quarter\lambda E(x)\sqrt{2c_-(x)}+\half c_-(x)-\quarter (Ax+B)\,,
\mea\mea
{\hat c}_-(x)=\quarter\lambda  E(x)\sqrt{2c_-(x)}+\half c_-(x)-\quarter (Ax+B)\,,
\mea\mea
{\hat E}(x)=\sqrt{2c_-(x)}\,/\lambda \,,\quad {\hat A}_+={\hat A}_-=-\quarter A\,,\quad {\hat B}=-\half B\,.
\label{gambier2}
\eea
\subsection{Inverse Gambier-transformed solutions}
These convert any solution ${\cal S}$ for which $A_+=A_-=A$,  
with a given value of $B$,  into  solutions with 
${\hat A}_+=0$ (resp. ${\hat A}_-=0$) and with ${\hat B}=-2B$.  In detail, ${\hat {\cal S}}={\cal G}_+^{-1}({\cal S})$ has                                                                                                                                                     \bea
{\hat c}_+(x)=\half\lambda^2 E(x)^2\,,\quad {\hat E}(x)=2(c_+(x)-c_-(x))/\lambda^2 E(x)\,,
\mea\mea
{\hat c}_-(x)=-\half\lambda^2  E(x)^2+2(c_+(x)-c_-(x))^2/\lambda^2E(x)^2 -4 Ax-2B\,,
\mea\mea
{\hat A}_+=0,\quad {\hat A}_-=-4  A\,,\quad {\hat B}=-2B\,.\qquad\qquad
\label{gambier3}
\eea
and similarly, ${\cal G}_-^{-1}({\cal S})$ has 
\bea
{\hat c}_-(x)=\half \lambda^2E(x)^2\,,\quad {\hat E}(x)=-2(c_+(x)-c_-(x))/\lambda^2E(x)\,,
\mea\mea
{\hat c}_+(x)=-\half \lambda^2 E(x)^2+2(c_+(x)-c_-(x))^2/\lambda^2 E(x)^2 -4 Ax-2B\,,
\mea\mea
{\hat A}_-=0,\quad {\hat A}_+=-4  A\,,\quad {\hat B}=-2B\,.\qquad\qquad
\label{gambier4}
\eea
The case when $E(x)\equiv 0$ needs special treatment (see Appendix A).
Note that 
\bea
{\cal G}_-^{-1}({\cal S})={\cal C}{\cal G}_+^{-1}({\cal S})\,.
\label{conjugate_gambier}
\eea 
\section{Sequences of solutions}
Repeated application of ${\cal B}$ and ${\cal B}^{-1}$ produces from {\em any} `seed' solution ${\cal S}$, 
a sequence ${\cal Q}_{\cal S}$ of solutions that, except in special cases (see below),
 is doubly infinite.   Thus
\bea
{\cal Q}_{\cal S}= \{\dots \,,{\cal B}^{-2}({\cal S})\,, {\cal B}^{-1}({\cal S})\,, 
{\cal S}\,, {\cal B}({\cal S})\,, {\cal B}^2({\cal S})\,,\dots \}
\label{B_sequence}
\eea
There is an associated conjugate sequence
\bea
{\cal Q}_{{\cal C}({\cal S})}&= &\{\dots \,,{\cal B}^{-2}{\cal C}({\cal S})\,, {\cal B}^{-1}{\cal C}({\cal S})\,, 
{\cal C}({\cal S})\,, {\cal B}{\cal C}({\cal S})\,, {\cal B}^2{\cal C}({\cal S})\,,\dots \}
\mea\mea
&=& {\cal C}\{\dots \,,{\cal B}^{2}({\cal S})\,, {\cal B}({\cal S})\,, 
{\cal S}\,, {\cal B}^{-1}({\cal S})\,, {\cal B}^{-2}({\cal S})\,,\dots \}
\mea\mea
&=& {\cal C}{\cal Q}_{\cal S}^T\,,
\label{CB_sequence}
\eea
where ${\cal Q}_{\cal S}^T$ is the transpose of the sequence ${\cal Q}_{\cal S}$, {\em i.e.} the sequence
obtained from ${\cal Q}_{\cal S}$ by replacing the $n$th member by the $(-n)$th member, for $n=0\,,\pm 1\,,\pm2\,,\dots$.  
There is also an associated reflected sequence 
\bea
{\cal Q}_{{\cal R}({\cal S})}&= &\{\dots \,,{\cal B}^{-2}{\cal R}({\cal S})\,, {\cal B}^{-1}{\cal R}({\cal S})\,, 
{\cal R}({\cal S})\,, {\cal B}{\cal R}({\cal S})\,, {\cal B}^2{\cal R}({\cal S})\,,\dots \}
\mea\mea
&=& {\cal R}\{\dots \,,{\cal B}^{-2}({\cal S})\,, {\cal B}^{-1}({\cal S})\,, 
{\cal S}\,, {\cal B}({\cal S})\,, {\cal B}^{2}({\cal S})\,,\dots \}
\mea\mea
&=& {\cal R}{\cal Q}_{\cal S}\,.
\label{RB_sequence}
\eea

Because the quantities $\theta$ and $P(x)-\theta x$ (with value $B$) are invariant under the action of ${\cal C}$, ${\cal B}$ and ${\cal B}^{-1}$,
every  solution in ${\cal Q}_{\cal S}$ and  
${\cal Q}_{{\cal C}({\cal S})}$ has the same values for these quantities.
On the other hand, 
in the reflected sequence
${\cal Q}_{{\cal R}({\cal S})}$, every solution has ${\widehat B}=B+\theta$ and 
${\widehat \theta}= -\theta$.
In this way,  each of the sequences ${\cal Q}_{\cal S}$, ${\cal Q}_{{\cal C}({\cal S})}$ and ${\cal Q}_{{\cal R}({\cal S})}$
is {\em partially} labelled by the values of $B$ and $\theta$.
\section{B\"acklund flux-quantization}
It follows from (\ref{backlund}) and (\ref{backlund_inverse}) that, if we write for the moment
\bea
{\cal B}^n({\cal S})&=&({ c}_+^{(n)}\,,{ c}_-^{(n)}\,,{ E}^{(n)}\,,{ A}_+^{(n)}\,,{ A}_-^{(n)})\,,\quad  
n\in\{0\,,\pm 1\,,\pm 2\,,\dots\}\,,
\label{Jpmvalues1}
\eea
then 
\bea
A_+^{(n)}+A_-^{(n)}=\theta\,,\qquad\qquad
\mea\mea
{ A}_+^{(n)} - { A}_-^{(n)} =A_+^{(0)} - A_-^{(0)} + 2n\theta\,.
\label{Jpmvalues2}
\eea
Here $\theta$ may take any real value. 
If $\theta\ne 0$, we see from (\ref{Jpmvalues2})  that without loss of generality we can assume  for ${\cal S}$ that 
\bea
A_+^{(0)} - A_-^{(0)} =\phi\,,\quad -\theta\leq \phi < \theta\,,
\label{Jpmvalues3}
\eea
and we then have
\bea
{ A}_+^{(n)}&=& (\half + n)\theta +\half \phi  \,,
\mea\mea
{ A}_-^{(n)}&=& (\half -n)\theta -\half\phi  \,,\quad n\in\{0\,,\pm 1\,,\pm 2\,,\dots\}\,.
\label{Jpmvalues4}
\eea
It follows that the steady ionic fluxes
as in \eqref{system1}  are also quantized 
in each of the sequences of solutions generated by B\"acklund transformations
and their inverses, with values for the $n$--th element of the sequence given by 
\bea
\Phi_+^{(n)} =(n+1)\Phi_+^{(0)} + n(D_+/D_-)\Phi_-^{(0)}\,,
\mea\mea
\Phi_-^{(n)}=-(n-1)\Phi_-^{(0)} -n (D_-/D_+)\Phi_+^{(0)}\,,
\label{flux_values}
\eea
and then the electric current density as in \eqref{current1} is  also quantized, with $n$--th value given in dimensionless
form by
\bea
j^{(n)}=[\alpha_-(\theta-\phi)-\alpha_+(\theta+\phi)]/2-n\theta(\alpha_+ + \alpha_-)
\label{dimensionless_current_quantization}
\eea
and in dimensional form by  
\bea 
J^{(n)}&=&({\hat z}e)\left[[n+1+n(D_-/D_+)]\Phi_+^{(0)} + [n-1+n(D_+/D_-)]\Phi_-^{(0)}\right]
\mea\mea
&=& J^{(0)}+ n\Delta J\,,\quad 
\Delta J={\tilde z}e(D_++D_-)\left\{ \frac{\Phi_+^{(0)}}{D_+}+\frac{\Phi_-^{(0)}}{D_-}\right\}\,.
\label{current_quantization}
\eea
The case  $\theta=0$ (zero resistive force-density) is degenerate; then $\phi$ may take any real value, and there is 
no flux-quantization, as 
\eqref{Jpmvalues4} shows.  In this case, the system \eqref{system2} is solvable explicitly in terms of elliptic functions, 
corresponding to the reduction of PII in \eqref{painleve1} to an elliptic ODE \cite{grafov,rogers1,whittaker}.   

Flux quantization is a remarkable feature of the structure of solutions of \eqref{system1} when grouped into 
sequences by B\"acklund transformations. 
To begin to explore the physical interpretation of this mathematical 
result, we reconsider Planck's electrically neutral solution \eqref{planck} in the case 
$E(x)\equiv 0$ when it becomes an exact solution of \eqref{system2}, that is, when $A_+=A_-=c_1-c_0=A$.  We take this as     
seed solution ${\cal S}$ in a doubly-infinite sequence ${\cal Q}_{\cal S}$ of exact solutions generated by B\"acklund transformations 
and their inverses.  Writing for the 
moment 
${\hat{\cal S}}={\cal B}({\cal S})$ and  ${\cal S}^\dag={\cal B}^{-1}({\cal S})$ for the two members of this sequence on either side of ${\cal S}$,   
we find using \eqref{backlund}
and \eqref{backlund_inverse} that ${\hat{\cal S}}$ has the form
\bea
{\hat c}_+(x)=c_0+Ax+2\lambda^2 A^2/(c_0+Ax)^2\,,\quad {\hat c}_-(x)=c_0+Ax\,,
\mea\mea
{\hat E}(x)=-2A/(c_0+Ax)\,,\quad {\hat A}_+=3A\,,\quad {\hat A}_-=-A\,;
\label{excited_up}
\eea
${\cal S}$ has the form (from \eqref{planck})
\bea
c_+(x)=c_-(x)=c_0+Ax\,,
\mea\mea
E(x)=0\,,\quad A_+=A_-=A\,;
\label{ground}
\eea
and ${\cal S}^\dag$ has the form 
\bea
c^{\dag}_-(x)=c_0+Ax+2\lambda^2A^2/(c_0+Ax)^2\,,\quad c^{\dag}_+(x)=c_0+Ax\,,
\mea\mea
E^{\dag}(x)=2A/(c_0+Ax)\,,\quad A^{\dag}_+=-A\,,\quad A^{\dag}=3A\,.
\label{excited_down}
\eea
We note that \eqref{excited_down} can also be obtained from \eqref{excited_up} using \eqref{CRB_relations}.  In each of these three solutions, the
concentrations are everywhere positive as required.  It is natural to think of ${\cal S}$ 
as describing a `ground state' of 
the system, where $E(x)\equiv 0$ and $j=(\alpha_--\alpha_+)A$.  Then ${\hat{\cal S}}$
and ${\cal S}^\dag$ describe `excited states,' with $j=-(3\alpha_++\alpha_-)A$ and $j=(\alpha_++3\alpha_-)A$ respectively.  
Note however that charge-neutrality at $x=0$ and $x=1$ is
satisfied by the seed solution ${\cal S}$, but not by ${\hat{\cal S}}$ and ${\cal S}^\dag$, and 
it is not clear what interpretation to give to the boundary-values satisfied in these two excited states.

The sequence ${\cal Q}_{\cal S}$ generated from ${\cal S}$ in this case corresponds to the well-known 
doubly-infinite sequence of rational 
solutions  of PII \cite{yablonskii,vorobev,clarkson,noumi,rogers1,rogers6}.  Expressions for the  members of the sequence beyond 
those given in \eqref{excited_up}-\eqref{excited_down} 
become increasingly more complicated,
but it is straightforward to use MATLAB \cite{matlab} to construct 
such members approximately without solving \eqref{system2} numerically, simply 
by proceeding from the seed solution using \eqref{backlund} 
and \eqref{backlund_inverse}.
\begin{figure}[ht]
\centering
\mbox{\subfigure{\includegraphics[width=2.5in]
{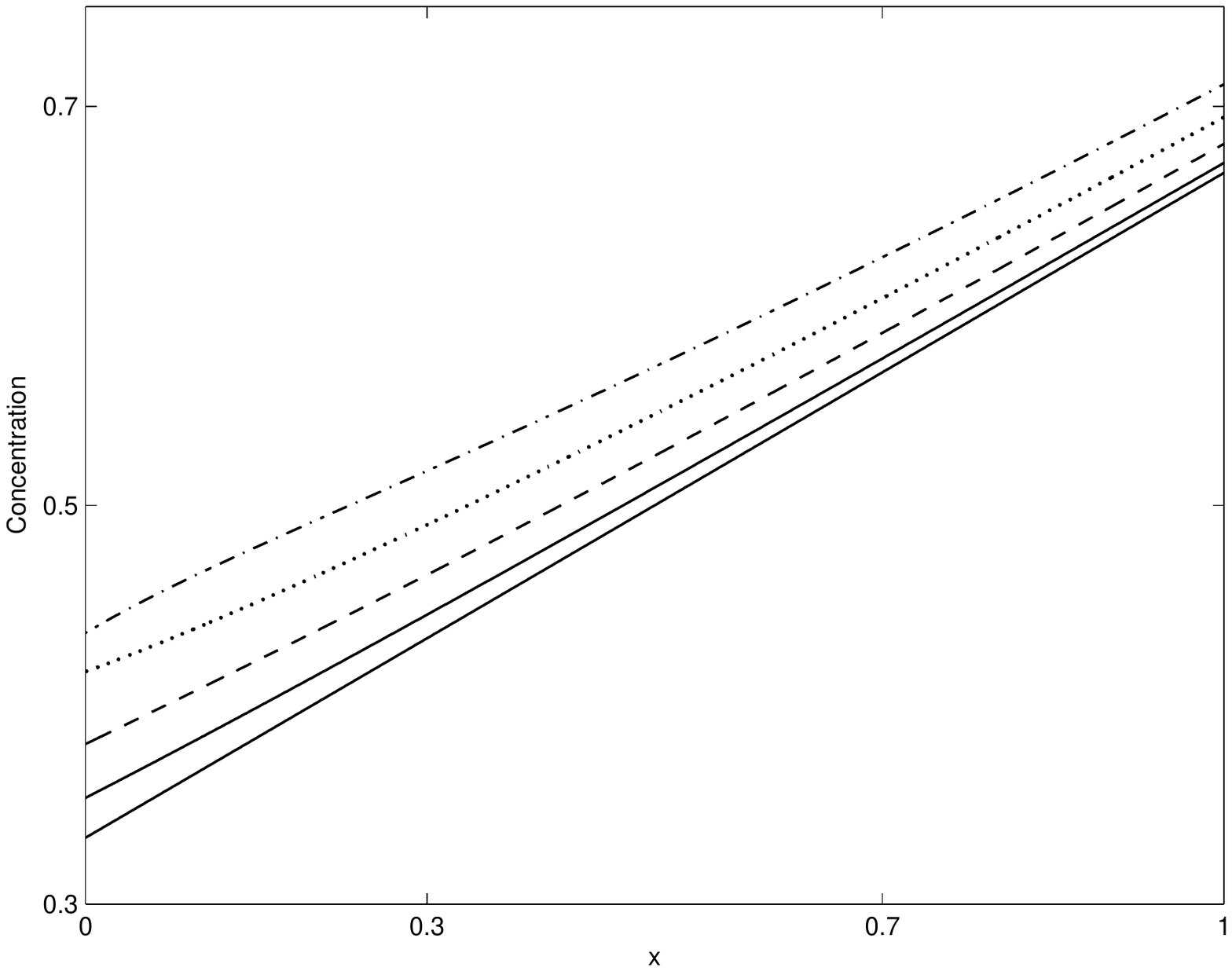}
}\quad
\subfigure{\includegraphics[width=2.5in]
{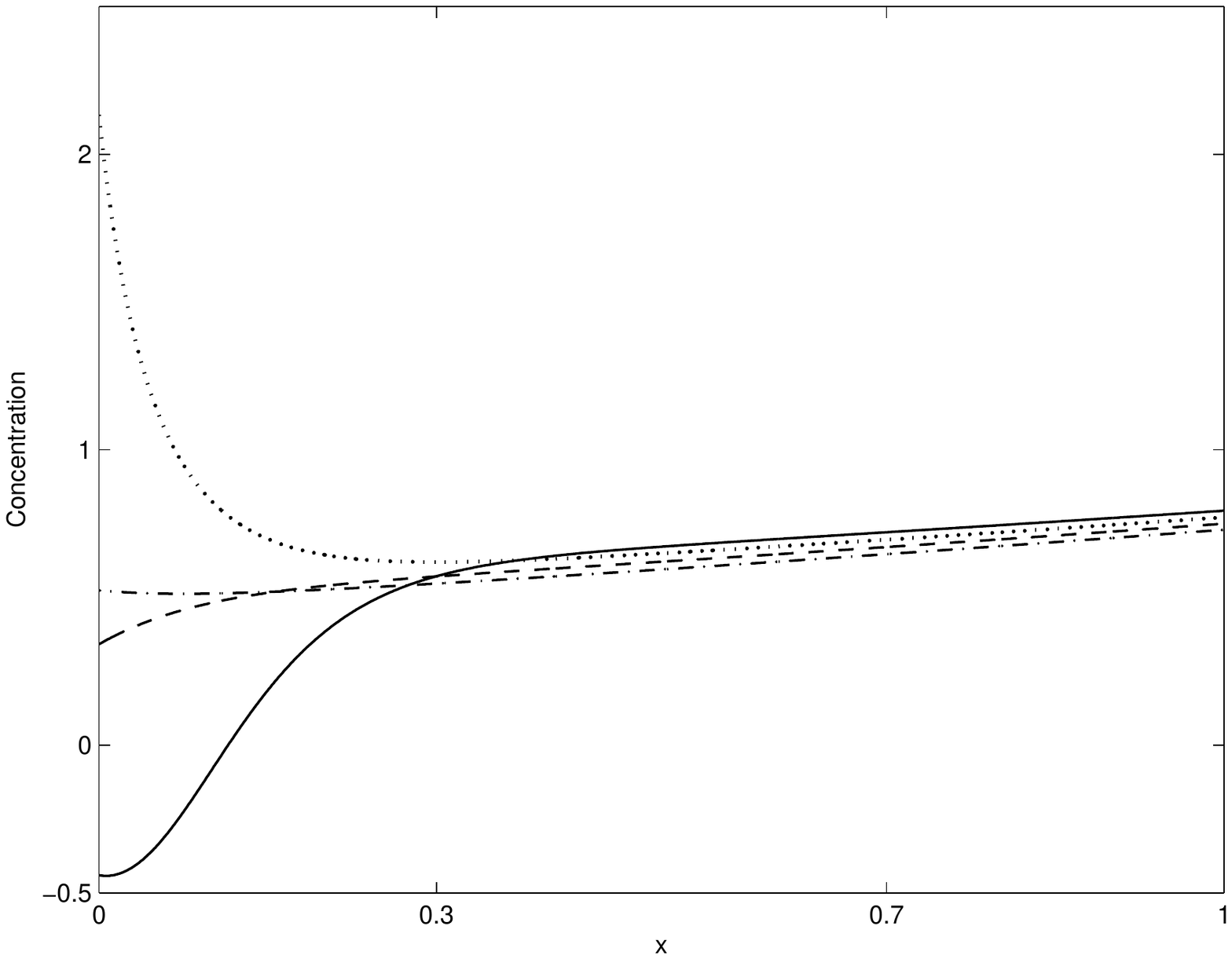} 
}}
\caption{On the left, graphs of $c_+(x)$ (straight, solid) as in \eqref{ground} and its $n$th transforms under ${\cal B}$, for $n=1$ (solid), $n=2$ (dashed), $n=3$ (dotted) and $n=4$ (dashdotted).  On the right, graphs for $n=5$ (dashdotted), $n=6$ (dashed), $n=7$ (dotted) and $n=8$ (solid), showing appearance of unphysical negative values in the final case.}
\end{figure}

Fig. 1 shows $c_+(x)$ as in \eqref{ground} and its first eight iterates under 
${\cal B}$, for the choices $c_0=1/3$,
$A=1/3$, and $\lambda^2=0.01$. 
Note from \eqref{backlund} that these are also the graphs of the first nine iterates under ${\cal B}$ 
of $c_-(x)$, and from \eqref{backlund_inverse}, they are also
the seed and first eight iterates under ${\cal B}^{-1}$ of $c_-(x)$, and also the first nine iterates under ${\cal B}^{-1}$ of $c_+(x)$.  
The ninth graph shows unphysical negative concentration values, 
showing that for these
parameter values, there is a ground state plus seven  excited states plus their seven conjugates.
The corresponding evenly-spaced current-density values, from the seventh iterate under ${\cal B}^{-1}$ through to the 
seventh iterate under ${\cal B}$ are, in dimensionless and dimensional form 
\bea
j^{(n)}&=&-[(2n+1)\alpha_+ + (2n-1)\alpha_-]A\,,
\mea\mea
J^{(n)}&=&{\hat z}e(c_0-c_1)[(D_+-D_-)+2n(D_+ + D_-)]/\delta\,,
\mea\mea
n &=&-7,\,-6,\,\cdots ,\,7\,.
\label{current_values}
\eea
Experiment with this technique using MATLAB shows that the number of excited states with positive concentrations obtained from \eqref{planck}
(in the case with $E(x)\equiv 0$), depends on the values of $c_0$, $c_1$ and $\lambda$.    
 
Examples like this show that for any physically sensible seed solution ${\cal S}$,  we must expect that only a finite subsequence of
${\cal Q}_{\cal S}$ may 
correspond to physically meaningful situations; in other words, only a finite number of physically acceptable solutions will be 
generated from  any acceptable  seed solution  by B\"acklund transformations and their inverses.  Within that subset of solutions, 
however, we have nevertheless  the remarkable result that ionic fluxes and 
the electric current-density are quantized in accordance with \eqref{flux_values}.

A notable feature of \eqref{current_quantization} and \eqref{current_values} is that increments of 
fluxes pertaining to successive values of $n$ 
are all equal and all determined at $n=0$, despite radical differences between the solutions of \eqref{system1} 
associated with those fluxes at different $n$ values.  What common element of such disparate solutions can engender 
consistency with the equality of the flux increments? We develop elsewhere \cite{bass4} the hypothesis that the common element 
is the quantum of electric charge.

[{\em Remark:} In regard to the general question as to when B\"acklund or inverse B\"acklund 
transformations of a given physically meaningful solution of \eqref{system2} give rise to 
new (potentially) physically meaningful solutions, with positive concentrations, 
we may make the following intuitive observations.  When $D_+\neq D_-$, it is clear that 
no steady electric field can counteract diffusion of both ionic species 
and force
both fluxes $\Phi_+$ and $\Phi_-$ to zero.  However, there is a field capable of equating the fluxes, so that positive and negative 
ions moving together make up zero electric current density; this is the situation where $J=0$ in \eqref{current1}.  
To achieve this, $E(x)$ must be directed 
so as to oppose the flux of the more mobile species.  Because $E(x)$ has to be anti-parallel (resp. parallel) to the 
flux of positive (resp. negative) ions in order to oppose it,  if $D_+>D_-$ then $E(x) \Phi_+ <0$, and if $D_+<D_-$ then $E(x)\Phi_->0$.
In dimensionless form, if $j=0$ and $\alpha_+>\alpha_-$ then $E(x)A_+>0$, and if  $j=0$ and $\alpha_+<\alpha_-$ then $E(x)A_-<0$.   
But then it is clear from \eqref{backlund} and \eqref{backlund_inverse}
that if $c_+(x)$ and $c_-(x)$ are positive, so are their B\"acklund transforms ${\hat c}_+(x)$
and ${\hat c}_-(x)$ in the first case, while their inverse B\"acklund transforms are positive in the second case.  
Note that this result is independent of the choice of  boundary conditions. 
For $j\neq 0$ the argument breaks down; and since a solution with $j=0$ is carried by a B\"acklund or inverse 
B\"acklund transformation into a solution with $j\neq 0$ in general, as \eqref{dimensionless_current_quantization} shows, 
the argument cannot be applied more than once to a solution with $j=0$; then 
we cannot easily tell if further B\"acklund or inverse B\"acklund transformations
will produce solutions with positive concentrations.]
 
It is clear that
an analysis of BCs and positivity 
requirements for solutions of \eqref{system2} is called for; we turn to this in the next section.   
\section{Boundary conditions and positivity of concentrations}
The ionic fluxes $\Phi_+$ and $\Phi_-$appearing in \eqref{system1}, and hence the dimensionless constants $A_+$ and $A_-$ 
appearing in \eqref{system2}, are not to be considered in general as given.  The electric current density \eqref{current1} and hence the linear 
combination \eqref{current2} may be prescribed,  but this leaves at least one of $A_+$ and $A_-$  to be determined as part of any 
solution ${\cal S}=(c_+\,,c_-\,,E\,,A_+\,,A_-)$ of \eqref{system2}.  Five pieces of data must be supplied to fix such a solution.
 
So far we have partially characterized the seed solution ${\cal S}$ of a sequence ${\cal Q}_{\cal S}$ by the values $B$, $\theta$ and $\phi$, thus 
fixing the values of $A_+$ and $A_-$.  Ignoring for the moment that this may not be possible in 
practice, we note that  
the conjugate
sequence ${\cal Q}_{{\cal C}({\cal S})}$ is then also partially characterized, by corresponding values $B$, $\theta$ and $-\phi$, and the reflected sequence ${\cal Q}_{{\cal R}({\cal S})}$ is also partially characterized, by corresponding values
$B-\theta$, $-\theta$ and $-\phi$.
Two further numbers are needed to fix each of these three sequences uniquely.  They can be taken to be the two 
further numbers that are needed to fix 
the seed solution ${\cal S}=(c_+\,,c_-\,,E\,,A_+\,,A_-)$, which so far has, using (\ref{first_integral})
and (\ref{Jpmvalues4}),
\bea
A_{+}=(\theta + \phi)/2\,,\quad A_{-}=(\theta - \phi)/2\,,\qquad\qquad\qquad
\mea\mea
c_+(0)+c_-(0)-\half\lambda^2 E(0)^2=B=c_+(1)+c_-(1)-\half \lambda^2 E(1)^2-\theta\,.
\label{seed_values1}
\eea
For example, we could prescribe the values at $x=0$ of all three of $c_+$, $c_-$ and $E$, bearing in mind that
we have already prescribed the value $B$
of the combination $c_+(0)+c_-(0)- \lambda^2 E(0)^2/2$.    
Then we can reasonably hope that the seed solution ${\cal S}$ is
uniquely determined, and so in turn, every element 
of the sequences ${\cal Q}_{\cal S}$, ${\cal Q}_{{\cal C}({\cal S})}$ and ${\cal Q}_{{\cal R}({\cal S})}$.

Writing again ${\cal B}^n({\cal S})=({\hat c}_+\,,{\hat c}_-\,,{\hat E}\,,{\hat A}_+\,,{\hat A}_-)$, 
we could then determine the values of ${\hat c}_+(0)$, ${\hat c}_-(0)$ and ${\hat E}(0)$ by repeated use of the formulas
(\ref{backlund}) and (\ref{backlund_inverse}) at $x=0$, although the resulting expressions 
quickly become very complicated as $|n|$ increases.  Similarly, if we were to fix ${\cal S}$ by prescribing 
the values of $c_+(1)$, $c_-(1)$ and $E(1)$, we could then determine the values of 
${\hat c}_+(1)$, ${\hat c}_-(1)$ and ${\hat E}(1)$ by repeated use of the formulas
(\ref{backlund}) and (\ref{backlund_inverse}) at $x=1$.  Note however that 
if we were to fix the seed solution by giving the values of, say,
$E(0)$ and $E(1)$ in addition to (\ref{seed_values1}), then we could not 
determine the values of $\widehat{E}(0)$ and $\widehat{E}(1)$ using (\ref{backlund}) and (\ref{backlund_inverse}) alone. 

Because the values of $A_+$ and $A_-$ are not in general known {\em a priori}, 
such considerations are of more mathematical than physical interest, and we now turn to more physically relevant ways of prescribing the 
five pieces of data needed to fix a solution of \eqref{system2}.  Two types of BCs are of particular interest.
\subsection{Charge-neutral boundary conditions}  
These were adopted by Planck \cite{planck} and in the formulation of the present model given in \cite{bass1, bass2}, and take
the form
\bea
c_+(0)=c_-(0)=c_0>0\,\,{\rm  and}\,\, c_+(1)=c_-(1)=c_1>0\,.
\label{neutralBCs}
\eea
Without loss of generality, we can suppose that $c_0\leq c_1$.  Taking $c_{ref.}$
to equal the sum of the concentrations at the two faces of the slab before proceeding to 
dimensionless variables as in \eqref{change_variables}, 
we then have in dimensionless form
\bea
0<c_0\leq c_1\,,\quad c_0+c_1=1\,.
\label{neutral1}
\eea
Eqs. \eqref{neutralBCs}  provide four of the five pieces of data that we expect are needed to determine
a solution of \eqref{system2}.  The remaining piece is typically provided by fixing the value of the current density,
\bea
j=j_0\,.
\label{currentBC}
\eea
Using the last of \eqref{system2}, we see from \eqref{neutralBCs} that
\bea
E\,'(0)=0=E\,'(1)\,,
\label{neumann1}
\eea
implying Neumann BCs for PII as in \eqref{painleve1}.  However this boundary-value problem  for \eqref{painleve1} is not of a standard form because
constants to be determined as part of  the solution appear in the coefficients of the ODE itself.  In full, 
we have using \eqref{first_integral}, \eqref{neutralBCs} and \eqref{currentBC} in \eqref{painleve1} that
\bea
\lambda^2\,E\,''(x)=\half\lambda^2E(x)^3\qquad\qquad\qquad\qquad\qquad\qquad\qquad\qquad\qquad\qquad
\mea\mea
\quad+\left[2c_0-\half\lambda^2\,E(0)^2+\left\{2(c_1-c_0)+\half\lambda^2(E(0)^2-E(1)^2)\right\}x\right]E(x)
\mea\mea
-(\alpha_+-\alpha_-)\left\{2(c_1-c_0)+\half\lambda^2(E(0)^2-E(1)^2)\right\}-2j_0\,.
\label{painleve_neumann}
\eea  
Here $\lambda$, $c_0$, $c_1$, $\alpha_+$, $\alpha_-$ and $j_0$ are known constants, while $E(0$ and $E(1)$ are to be determined
together with $E(x)$, for $0<x<1$.   
Despite its unusual form, in the case $j_0=0$ existence of positive, strictly decreasing solutions $E(x)$ of \eqref{painleve_neumann} and \eqref{neumann1} satisfying
$c_0E(0)<c_1E(1)$, and non-existence of negative solutions, 
has been shown \cite{thompson, amster} for $\alpha_+>\alpha_-$.  
By charge-conjugation, existence of negative, strictly increasing solutions
satisfying $c_0E(0)>c_1E(1)$, and non-existence of positive solutions, 
has been shown for $\alpha_+<\alpha_-$.  
If $\alpha_+=\alpha_-$, then $j_0=0$ implies that $A_+=A_-=0$, and  the explicit solution $E(x)\equiv 0$  exists 
as in \eqref{planck}.
 
From any of these solutions $E(x)$, a solution to \eqref{system2}, \eqref{neutralBCs} and \eqref{currentBC} can
be constructed using \eqref{first_integral} and the last of \eqref{system2}.   
Note that when $E(x)$ is strictly increasing (resp. decreasing), then $c_+(x)>c_-(x)$ (resp. $c_+(x)<c_-(x)$) on $(0,1)$. 

Uniqueness has not been established for any  of these solutions, and there are no general existence or uniqueness results when $j_0\neq 0$.  

It has proved difficult \cite{amster} to solve \eqref{painleve_neumann} directly by numerical methods in order to obtain 
approximate solutions that illustrate the results obtained analytically \cite{thompson,amster}.  This difficulty, which stems from the fact that
the ODE to be solved involves constants that have to be determined as part of the solution, 
can be circumvented simply, by working with the system
\eqref{system2} rather than its consequence \eqref{painleve_neumann}, or more precisely, by rewriting \eqref{system2} as a system of five coupled
first-order ODEs,  
\bea
c_+\,'(x)= E(x)\,c_+(x)+A_+(x)\,,\quad
c_-\,'(x)= -E(x)\,c_-(x)+A_-(x)\,,
\mea\mea
\lambda^2 E\,'(x)=c_+(x)-c_-(x)\,,\quad
A_+\,'(x)=0\,,\quad A_-\,'(x)=0\,,\qquad
\label{system3}
\eea 
which does not involve any unknown constants.  
The MAPLE \cite{maple} routine {\em dsolve} solves \eqref{system3} almost instantaneously when supplemented by BCs of the form
\eqref{neutralBCs} and \eqref{currentBC}, with the latter rewritten as
\bea
\alpha_- A_-(0)-\alpha_+ A_+(0)=j_0\,.
\label{currentBC2}
\eea
Fig. 2 shows plots obtained in this way, for  $c_0=1/3$, $c_1=2/3$, $\lambda=0.5$ and $j_0=0$,
showing expected behaviour \cite{thompson,amster}: a positive, monotonically decreasing $E(x)$ on the left, in the case 
$\alpha_+=0.8>\alpha_-=0.2$, and a 
negative, monotonically increasing $E(x)$ on the right in the case $\alpha_+=0.4<\alpha_-=0.6$.  
On the left, $c_0E(0)\approx 0.15<c_1E(1)\approx 0.25$; on the right, $c_0E(0)\approx -0.049>c_1E(1)\approx -0.083$. 
In each case, \eqref{neumann1} is satisfied.   
\begin{figure}[!ht]
\centering
\mbox{\subfigure{\includegraphics[width=2.5in]
{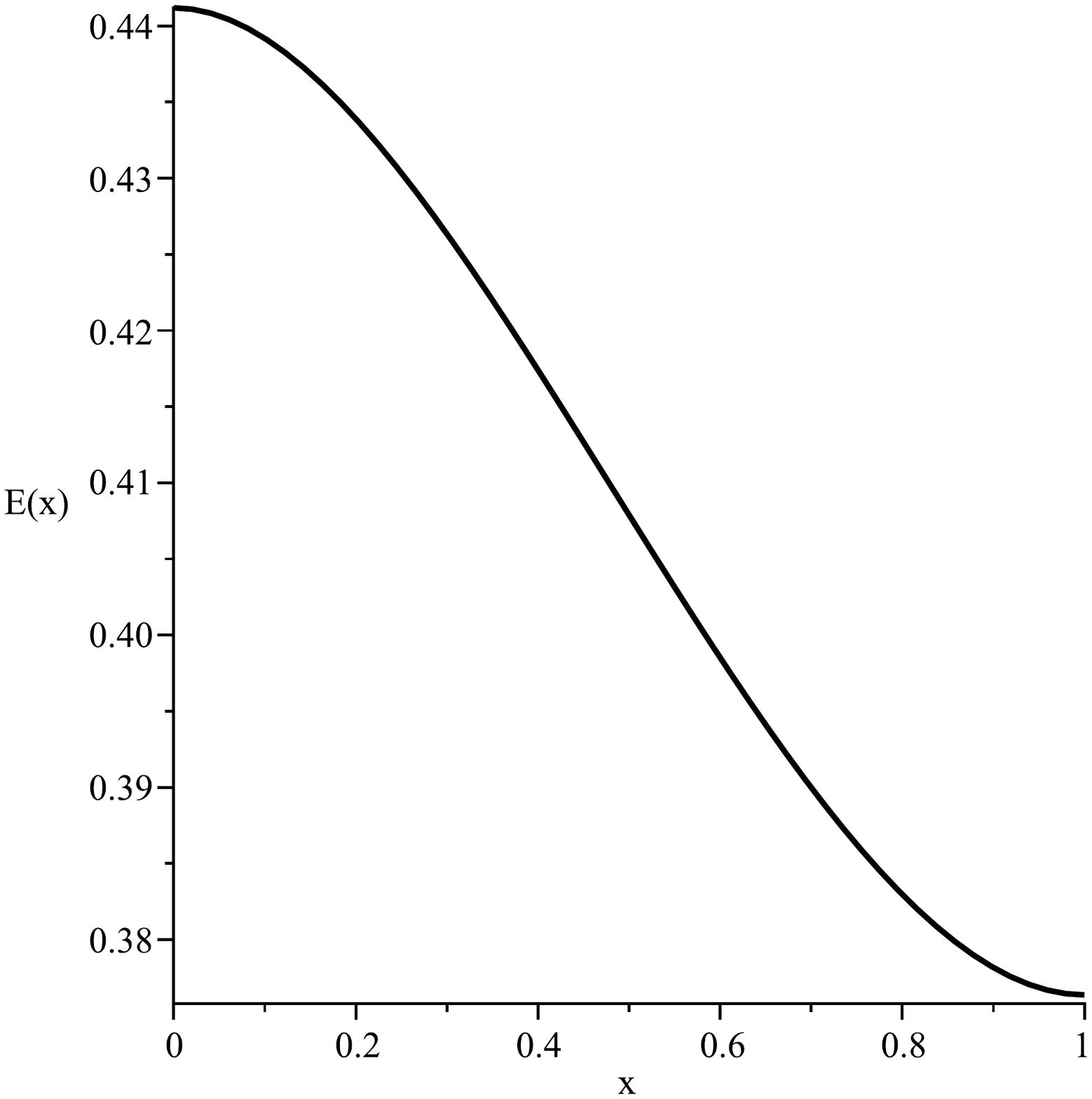}
}\quad
\subfigure{\includegraphics[width=2.5in]
{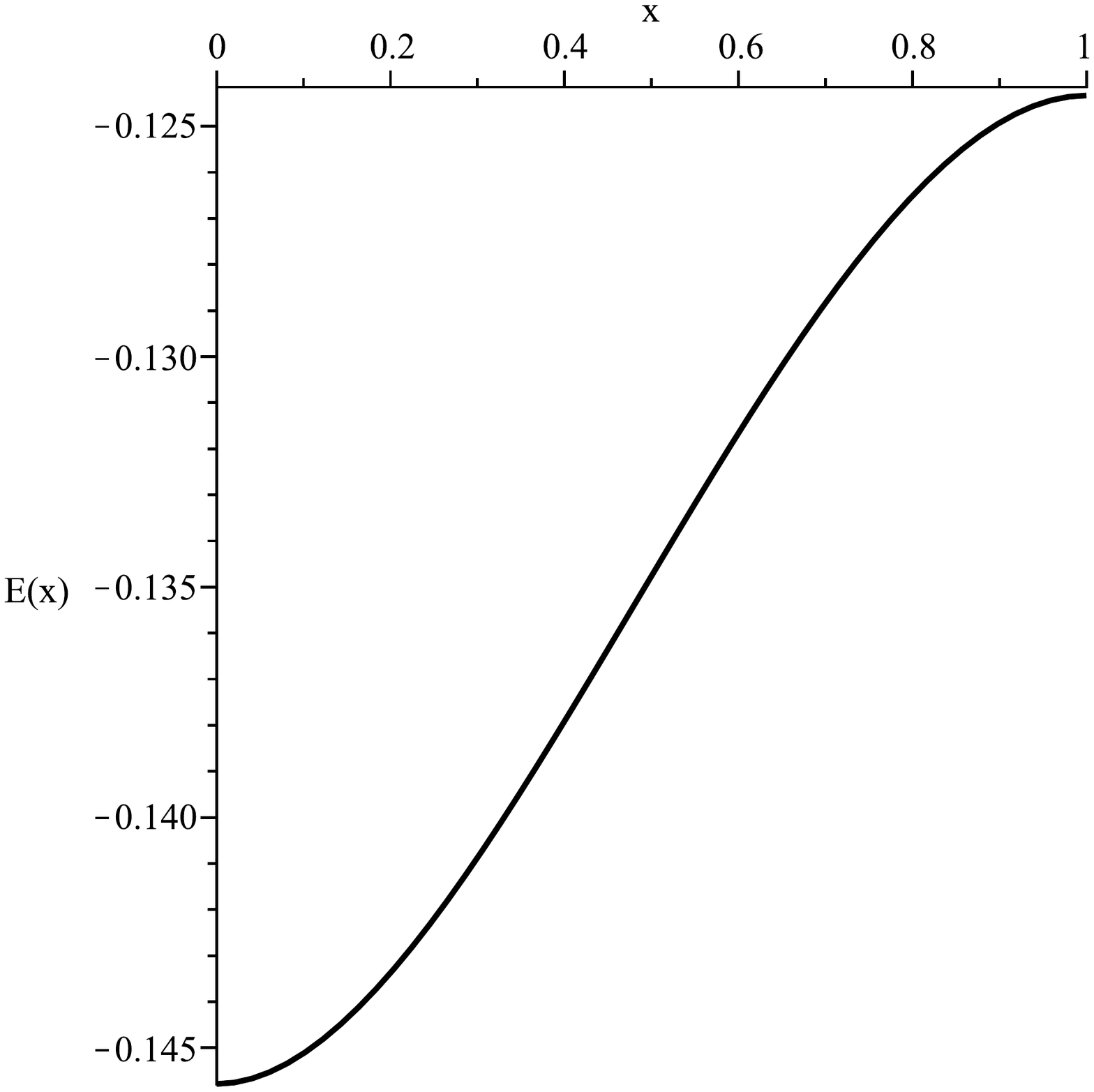} 
}}
\caption{Graphs of $E(x)$ when $c_0=1/3$, $c_1=2/3$, $j_0=0$ and $\lambda=0.5$. On the left, $\alpha_+=0.8$, $\alpha_-=0.2$; on the right,
$\alpha_+=0.4$, $\alpha_-=0.6$.}
\end{figure}

Comparison with the behaviour of $E(x)$ in cases with $j_0\neq 0$, as in Fig. 3 below, suggests strongly that these known existence and 
monotonicity results \cite{thompson,amster} can be extended to such cases also.  
  
It is easy to show as follows that for any solution of \eqref{system2}, \eqref{neutralBCs} in which $c_+(x)$ and $c_-(x)$ are continuously 
once-differentiable
on $(0,1)$, and continuous from the right (resp. from the left) at $x=0$ (resp. $x=1$), then
both 
$c_+$ and $c_-$ are everywhere positive on $[0,1]$.  

Suppose  firstly that $c_+(x)< 0$ 
on some open subinterval of $[0,1]$.  Because $c_+(0)=c_0>0$, and $c_+(1)=c_1>0$, it then follows that 
there is at least one point $x_1$ and 
one point $x_2$ in $[0,1]$ with 
\bea
c_+(x_1)=0\,,\quad c_+\,'(x_1)\leq 0\,,\quad c_+(x_2)=0\,,\quad c_+\,'(x_2)\geq 0\,.
\label{positivity1}
\eea
These conditions are only compatible with the first of (\ref{system2}) if 
\bea
c_+\,'(x_1)=c_+\,'(x_2)=A_+=0\,.
\label{positivity2}
\eea
But if $A_+=0$, it follows from the first of (\ref{system2}) and \eqref{neutralBCs} that
\bea
c_+(x)=c_0\,e^{\int_0^x\,E(y)\,dy}\,,\quad 0\leq x\leq 1\,,
\label{positivity3}
\eea
which is  everywhere positive, hence providing a contradiction. 
 
Next suppose instead that $c_+(x_3)=0$ at some point $x_3\in [0,1]$.  Because it cannot happen that $c_+(x)<0$ on an open subinterval of $[0,1]$, 
it follows that $x_3$ is a minimum, and $c_+\,'(x_3)=0$, so again from
the first of  (\ref{system2}) 
we have $A_+=0$, 
hence (\ref{positivity3}), and again a contradiction.

Thus $c_+>0$ on $[0,1]$, and similarly $c_->0$ on $[0,1]$. 

With that established, it is  possible to prove for any solution of this type that at least one of $A_+$ and $A_-$ must be positive if $c_1>c_0$.  
Assume to the contrary that $A_+\leq 0$ and $A_-\leq 0$.  From \eqref{system2} it then follows that
\bea
d\left( e^{\psi(x)}\,c_+(x)\right)/dx=A_+\,e^{\psi(x)}\leq 0\,,\mea\mea
d\left( e^{-\psi(x)}\,c_-(x)\right)/dx=A_-\,e^{-\psi(x)}\leq 0\,,
\label{Aproof1}
\eea
where
\bea
\psi(x)=-\int_0^x E(y)\,dy\,.
\label{Aproof2}
\eea
Then 
\bea
e^{\psi(0)}\,c_+(0)\geq e^{\psi(1)}\,c_+(1)\Rightarrow e^{\psi(0)-\psi(1)}\geq c_1/c_0\,,
\mea\mea
e^{-\psi(0)}\,c_-(0)\geq e^{-\psi(1)}\,c_-(1)\Rightarrow e^{\psi(0)-\psi(1)}\leq c_0/c_1\,,
\label{Aproof3}
\eea
which are inconsistent with $c_1>c_0$.

\begin{figure}[!]
\centering
\mbox{\subfigure{\includegraphics[width=2.5in]
{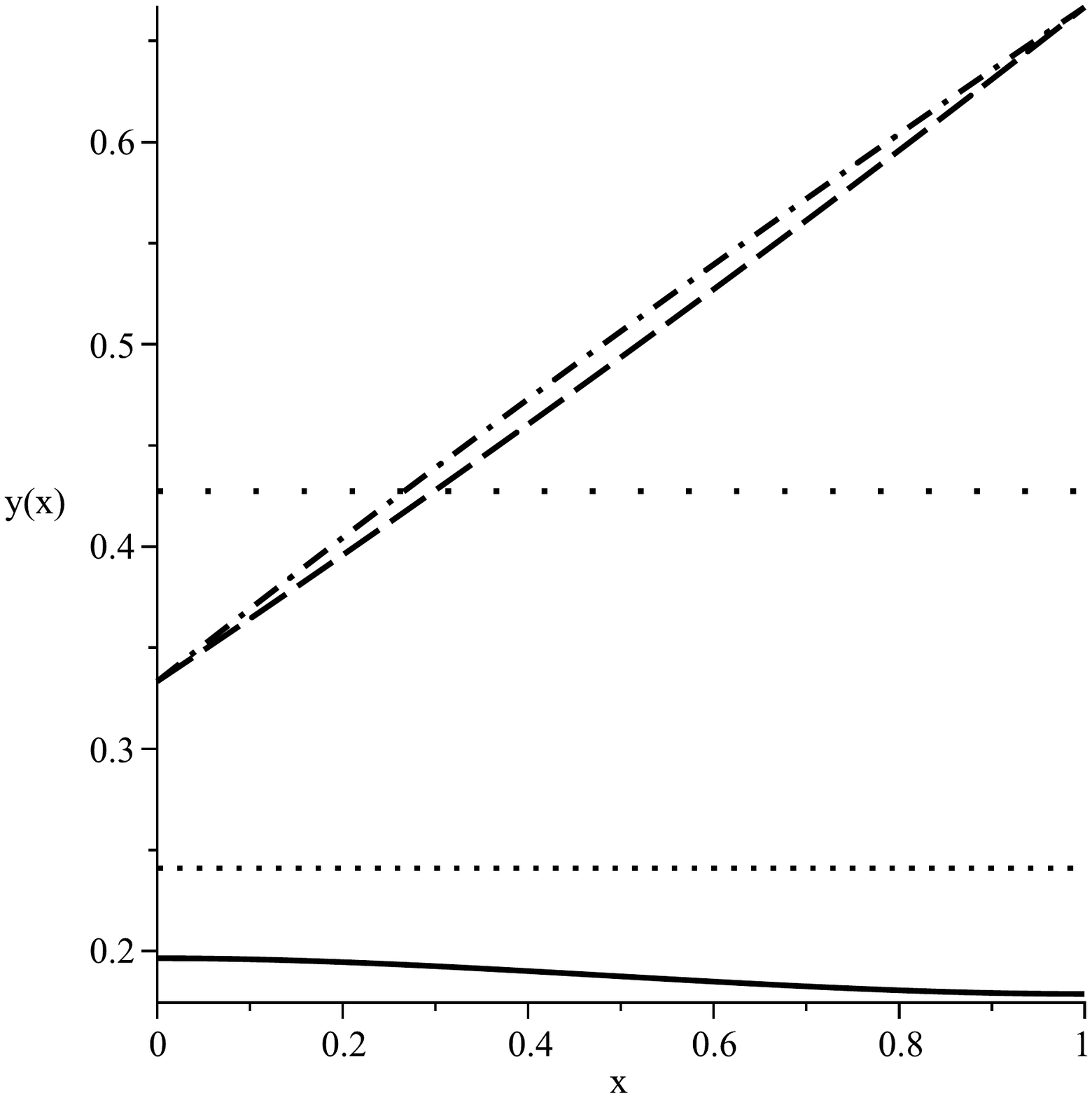}
}\quad
\subfigure{\includegraphics[width=2.5in]
{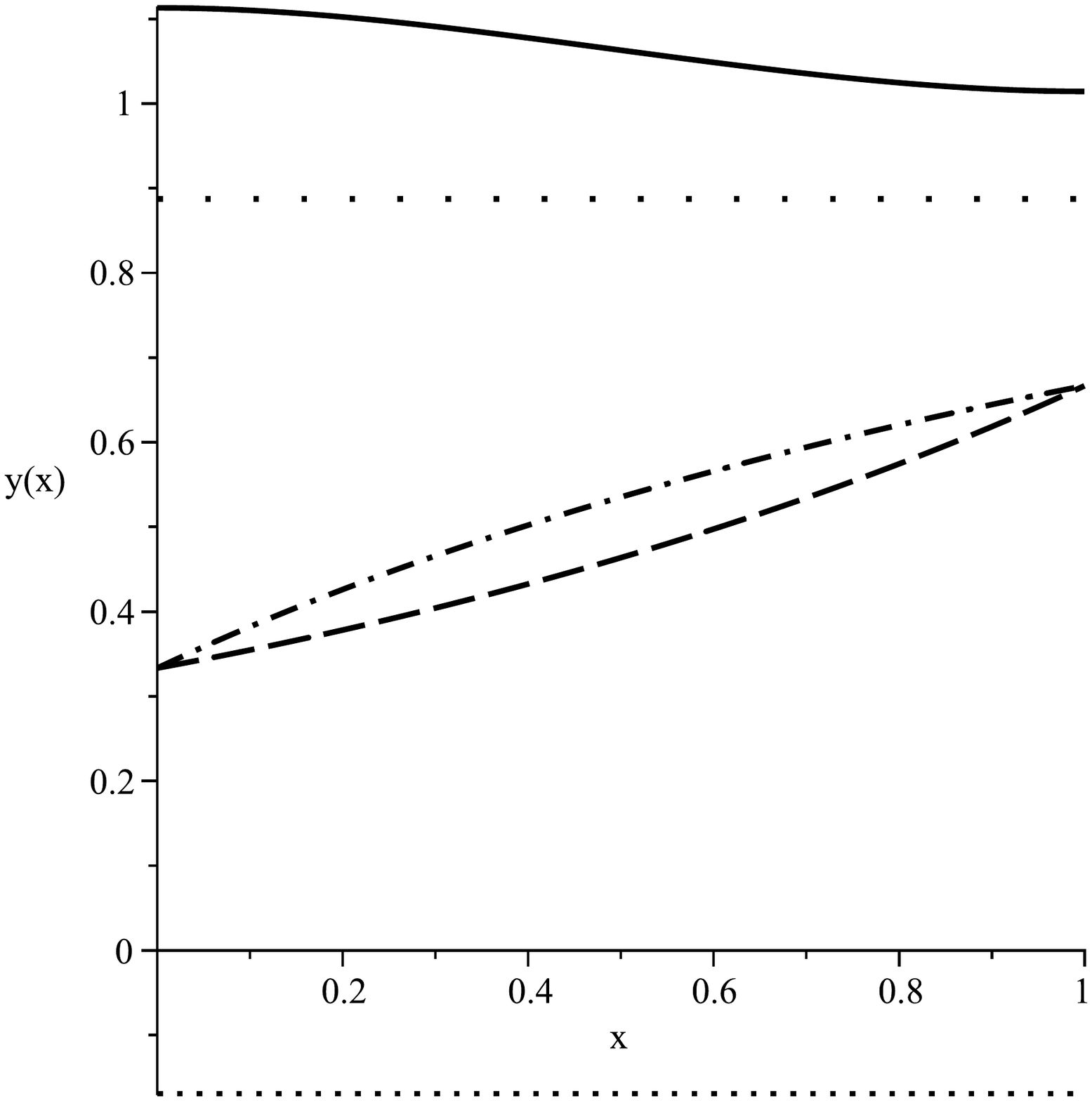} 
}}
\caption{Graphs of $c_+(x)$ (dash), $c_-(x)$ (dashdot), $E(x)$ (solid), $A_+$ (dot) and $A_-$ (spacedot), 
corresponding to charge-neutral BCs \eqref{neutralBCs}, \eqref{currentBC}, with $c_0=1/3$, $c_1=2/3$, $\lambda=0.7$, $\alpha_+=0.4$, $\alpha_-=0.6$ 
in each case, and 
with $j_0=0.16$ on the left, and $j_0=0.6$ on the right.
Both $A_+$ and $A_-$ are positive on the left; on the right, $A_-$ is positive but $A_+$ is negative.  }
\end{figure}

Numerical experiments indicate that there are cases for which both of  $A_+$, $A_-$ are positive, 
and cases where only one is positive.   Fig. 3, obtained in the same way as Fig. 2 above,  shows examples of both behaviours.     
Note that if $j_0=0$ in \eqref{currentBC}, then \eqref{current3} implies that
$A_+$ and $A_-$ have the same sign, and so both must be positive if $c_1>c_0$.

We now show that  BCs of the form \eqref{neutralBCs} and \eqref{neutral1} 
are not preserved under B\"acklund transformations.  We have already seen this in the special case of the exact Planck solution,
described in \eqref{planck}. Consider ${\hat {\cal S}}={\cal B}({\cal S})$ where ${\cal S}$ is a solution 
satisfying \eqref{neutralBCs} 
for some $c_0$ and $c_1$ as in 
\eqref{neutral1}. From \eqref{backlund} we see that ${\hat c}_+(0)={\hat c}_-(0)$ requires that
\bea
2A_+\lambda^2 (c_0E(0)+A_+)/c_0^2=0\,,
\label{nogo1}
\eea
which in turn requires either $A_+=0$ or $c_+\,'(0)\,\, (= c_0E(0)+A_+)=0$.  Similarly, either $A_+=0$ or $c_+\,'(1)=0$.  
But since ${\cal S}$ should be fixed by the 
five pieces of data in 
\eqref{neutralBCs} and \eqref{currentBC}, it is not possible to 
{\em impose} any of these conditions, and none will hold in general.    
Certainly there can be solutions ${\cal S}$ for which \eqref{neutralBCs} hold together with $A_+=0$ rather than \eqref{currentBC}, 
but then the corresponding current-density $j$ is not known {\em a priori}, but is determined as part of the solution.  
In such a case, ${\cal B}({\cal S})$ reduces to ${\cal C}({\cal S})$, the charge-conjugate of ${\cal S}$, 
with ${\hat A}_+=A_-$ and ${\hat A}_-=0$, as noted after \eqref{backlund}.  
It follows that a second iteration of ${\cal B}$ is not possible unless it is also true that $A_-=0$; but when $A_+=A_-=0$,   
no flux-quantization occurs, as \eqref{flux_values} shows.  
Corresponding remarks apply for inverse B\"acklund transformations. 
Thus charge-neutral BCs \eqref{neutralBCs} 
are not preserved in form under flux-quantization by B\"acklund transformations and their inverses.

\subsection{Corrected boundary conditions}
In proposing charge-neutral BCs \eqref{neutralBCs}, we have in mind  reservoirs of well-stirred ionic solutions occupying 
$-\infty<x<0$ and $1<x<\infty$, with $c_+(x)$ and $c_-(x)$ both equal to $c_0$ throughout the left-hand reservoir, 
and both equal to $c_1$ throughout the right-hand reservoir, and with $E(x)=0$ throughout both reservoirs. Stirring leads to the
diffusion coefficients $D_+$ and $D_-$  becoming effectively infinite, so that we also have 
$A_+=A_-=j=0$ throughout both reservoirs ({\em cf.} \eqref{system1} and \eqref{system2}).   
When, as we expect, \eqref{neutralBCs}  and \eqref{currentBC} determine a solution of \eqref{system2} in $0<x<1$,
boundary-values $E(0_+)$ and $E(1_-)$ of $E(x)$ are  determined in turn as part of that solution.  There is no reason to
expect that these values 
will be zero in general, and so they will not equal the assumed values in the reservoirs, $E(0_-)=E(1_+)=0$.  
But such jumps in the values of $E(x)$ across
the faces of the slab imply the presence of nonzero surface charges, contradicting the assumption of charge neutrality there.  
From this point of view, the   
charge-neutral BCs \eqref{neutralBCs} are inconsistent.

In reality, we must expect that the ionic concentrations and the electric field are continuous across each face of the slab.  
Then, for the left-hand reservoir for example, $c_+(0)\neq c_-(0)$ and $E(0)\neq 0$, but as $x\to -\infty$, 
$c_+(x)$ and $c_-(x)$ decay  to a common positive value
$c_{(-\infty)}$, while $E(x)$ decays to zero.  Similarly, in the right-hand reservoir as $x\to \infty$, $c_+(x)$ and 
$c_-(x)$ decay  to a common positive value
$c_{(+\infty)}$, while $E(x)$ decays to zero.  Without loss of generality, we can suppose that $c_{(+\infty)}\geq c_{(-\infty)}$.  
It is now appropriate to choose the sum  of 
these two 
limiting concentrations, before the change to dimensionless variables, as the $c_{ref}$ in \eqref{change_variables} and \eqref{change_constants}; 
in dimensionless terms we then have in place of \eqref{neutral1} that
\bea
0<c_{(-\infty)}\leq c_{(+\infty)}\,,\quad c_{(-\infty)}+c_{(+\infty)}=1\,.
\label{corrected1}
\eea

In the left-hand reservoir, equations of the form \eqref{system1} apply in simplified form, as $D_+\to\infty$ 
and $D_-\to\infty$; in dimensionless form, 
we get equations of the form \eqref{system2} with $A_+=A_-=0$, so that for $-\infty<x<0$ we have to solve
\bea
c_+\,'(x)=E(x)c_+(x)\,,\quad c_-\,'(x)=-E(x)c_-(x)\,,
\mea\mea
\lambda^2 E\,'(x)=c_+(x)-c_-(x)\,,\qquad\qquad\qquad
\label{reservoir1}
\eea
with $c_{\pm}(x)\to c_{(-\infty)}$ and $E(x)\to 0$ as $x\to -\infty$. Then the first two equations integrate to
\bea
c_+(x)= c_{(-\infty)}\,e^{-\varphi(x)}\,,\quad c_-(x)=c_{(-\infty)}\,e^{\varphi(x)}\,,
\label{reservoir2}
\eea
where
\bea
\varphi(x)=-\int_{-\infty}^x E(x)\,dx
\label{reservoir3}
\eea
is the electrostatic potential of the electric field in the reservoir.  
Because $E(x)=-\varphi '(x)$, the third of \eqref{reservoir1} then gives  the nonlinear Poisson-Boltzmann equation 
\bea 
-\lambda^2 \varphi ''(x)= c_{(-\infty)}\,\left[e^{-\varphi(x)}- e^{\varphi(x)}\right]\,,
\label{reservoir4}
\eea
to be solved on $-\infty<x<0$ subject to $\varphi(x)\to 0$ and $\varphi'(x)\to 0$ as $x\to -\infty$. 
The exact solution of this problem is known \cite{herzfeld,langmuir}.  In Appendix B we present this solution and describe the highly 
nonlinear exact BCs at the junction face(s) to which it leads. 
Here we suppose that
\bea
|\varphi(x)|\ll 1\,,
\label{reservoir5}
\eea
so that \eqref{reservoir4} linearizes to 
\bea
\lambda_0^2\, \varphi ''(x)=\varphi (x)\,,\quad \lambda_0=\lambda/\sqrt{2c_{(-\infty)}}\,.
\label{reservoir6}
\eea
with solution matching the limiting BCs at $x=-\infty$ given by
\bea
\varphi(x)=\varphi(0)\,e^{x/\lambda_0}\,.
\label{reservoir7}
\eea
In this approximation we then have, from \eqref{reservoir2} and \eqref{reservoir3},  
\bea
c_+(x)=c_{(-\infty)}\left[1-\varphi(0)\,e^{x/\lambda_0}\right]\,,
\mea\mea
c_-(x)=c_{(-\infty)}\left[1+\varphi(0)\,e^{x/\lambda_0}\right]\,,
\mea\mea
E(x)=-\frac{\varphi(0)}{\lambda_0}\,e^{x/\lambda_0}\,,
\label{reservoir8}
\eea
showing that $\lambda_0$ (in dimensional form, $\lambda_0 \delta$) can be interpreted as a 
Debye shielding length for the reservoir-junction interface.  It 
is the characteristic length over which $c_+(x)$, $c_-(x)$ and $E(x)$  decay to their limiting values as $x$ 
decreases from the interface value $x=0$.   In terms of the original variables, 
\bea
\lambda_0\delta= \sqrt{\epsilon k_B T/8\pi ({\tilde z}e)^2 c_{(-\infty)}}\,. 
\label{reservoir9}
\eea

[{\em Remark:} Just as $\lambda_0$ is a Debye length at concentration $c_{-\infty}$, so
the constant $\lambda/\sqrt{2}$ of \eqref{change_constants} is a Debye length at concentration $c_{ref.}$.  
In analyzing mathematically situations where $\lambda$ approaches zero \cite{bass5} (see Sec. 3), a different choice of $c_{ref.}$ in \eqref{change_constants}
may be more appropriate 
than the one we have made at the start of this section.  For example the choice of the 
logarithmic mean $(c_{(+\infty)}-c_{(-\infty)})/\ln(c_{(+\infty)}/c_{(-\infty)})$  as $c_{ref.}$ has the advantage that 
$c_{ref.}$ then tends to zero like the smaller of $c_{(-\infty)}$ and $c_{(+\infty)}$, guaranteeing that as $\lambda$ approaches zero,  
the Debye length in each reservoir also approaches zero.]

In \eqref{reservoir8}, the value of $\varphi(0)$ is determined from the solution of \eqref{system2} in the slab, 
which fixes $c_{\pm}(0)$ and $E(0)$.   
 
From \eqref{reservoir8} we have
\bea
c_+(0)+c_-(0)=2c_{(-\infty)}\,,\qquad\qquad\qquad
\mea\mea
c_+(0)-c_-(0)=-2c_{(-\infty)}\varphi(0)=2c_{(-\infty)}\lambda_0 E(0)\,,
\label{reservoir10}
\eea
and then, from the third of equations \eqref{reservoir1},
\bea
\lambda_0 E\,'(0)= E(0)\,.
\label{reservoir11}
\eea 

Similar considerations in the right-hand reservoir lead to
\bea
c_+(1)+c_-(1)=2c_{(+\infty)}\,,
\mea\mea
c_+(1)-c_-(1)=2c_{(+\infty)}\lambda_1 E(1)\,,
\label{reservoir12}
\eea
and
\bea
\lambda_1 E\,'(1)= E(1)\,.
\label{reservoir13}
\eea
In \eqref{reservoir12} and \eqref{reservoir13}, $\lambda_1$ is defined like $\lambda_0$ in \eqref{reservoir6}, with $c_{(+\infty)}$ replacing
$c_{(-\infty)}$.  

With $c_{(-\infty)}$, $c_{(+\infty)}$ and $j_0$ given constants, equations \eqref{reservoir10}, \eqref{reservoir12} and \eqref{currentBC} 
provide the new BCs that are to be applied
to fix a solution of the system \eqref{system2} on $0<x<1$, replacing \eqref{neutralBCs} and \eqref{currentBC}. Fig. 4 provides an 
illustration of the solution to the combined problem (left reservoir, slab, right reservoir).  In the slab, equations \eqref{system2} were solved
numerically using MAPLE, with the BCs \eqref{reservoir10} and \eqref{reservoir12}. 

\begin{figure}[!h]
\centering
\mbox{\subfigure{\includegraphics[width=1.63in]
{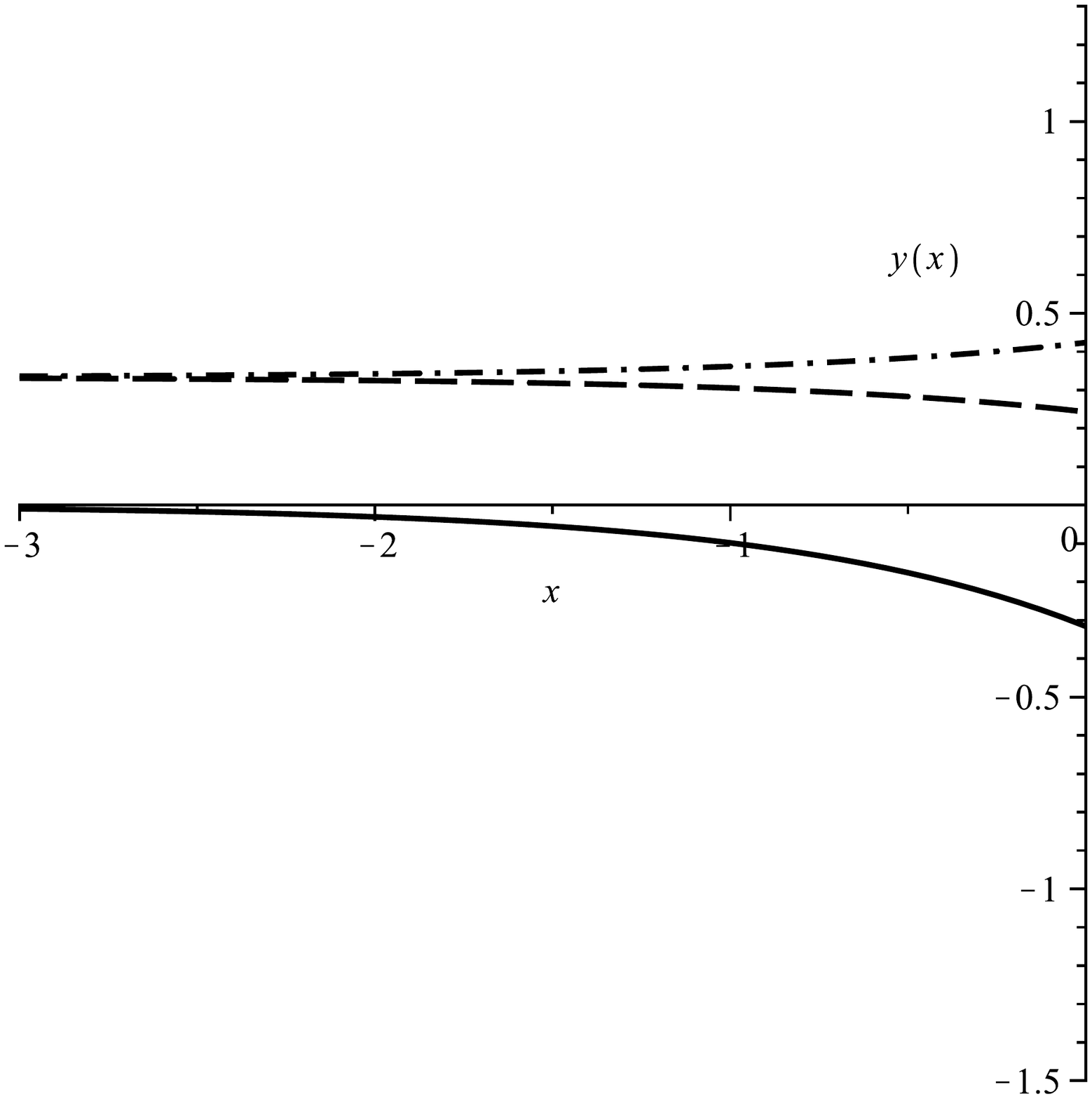}
}\quad
\subfigure{\includegraphics[width=1.63in]
{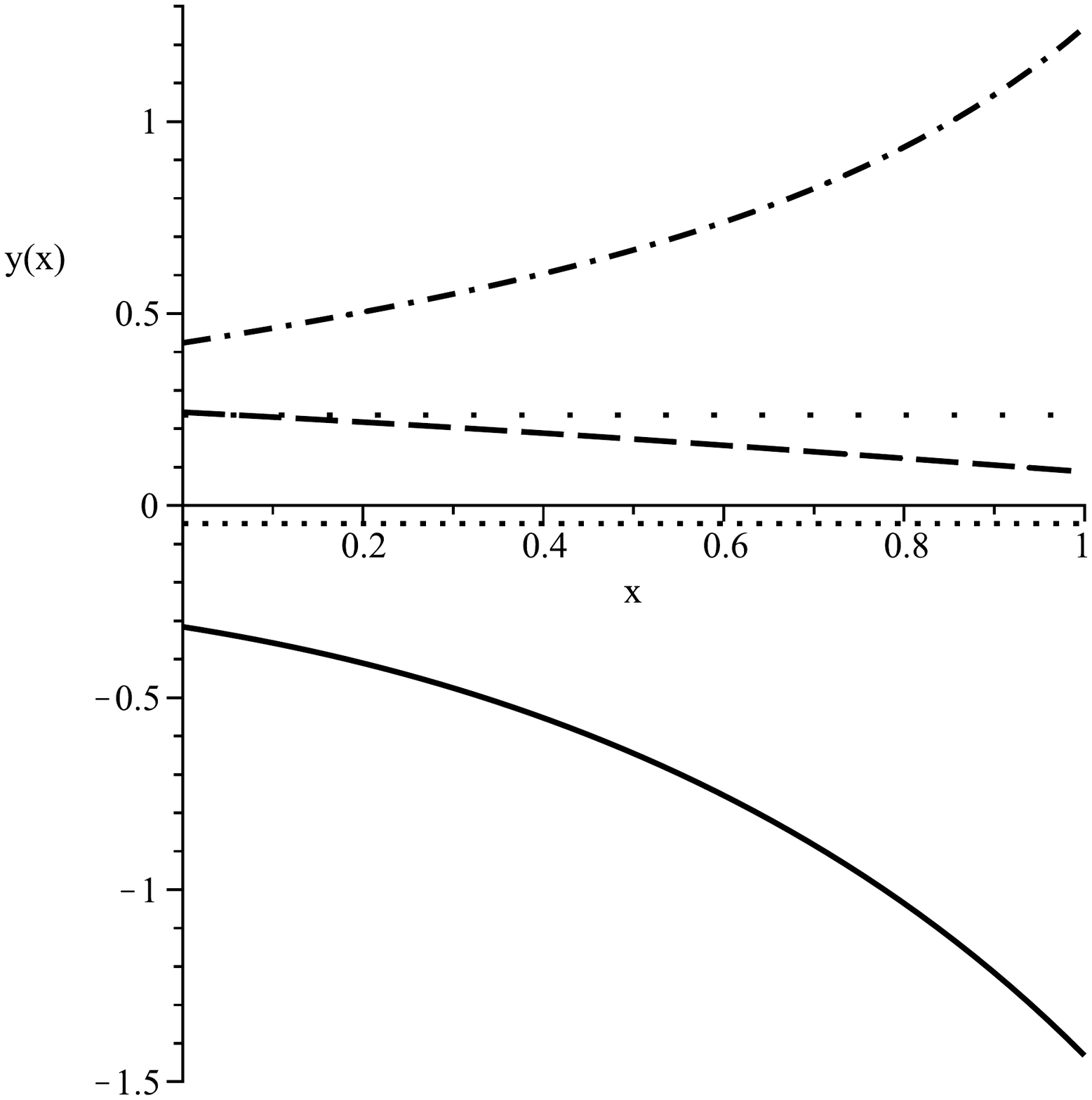}
 }\quad
\subfigure{\includegraphics[width=1.63in]
{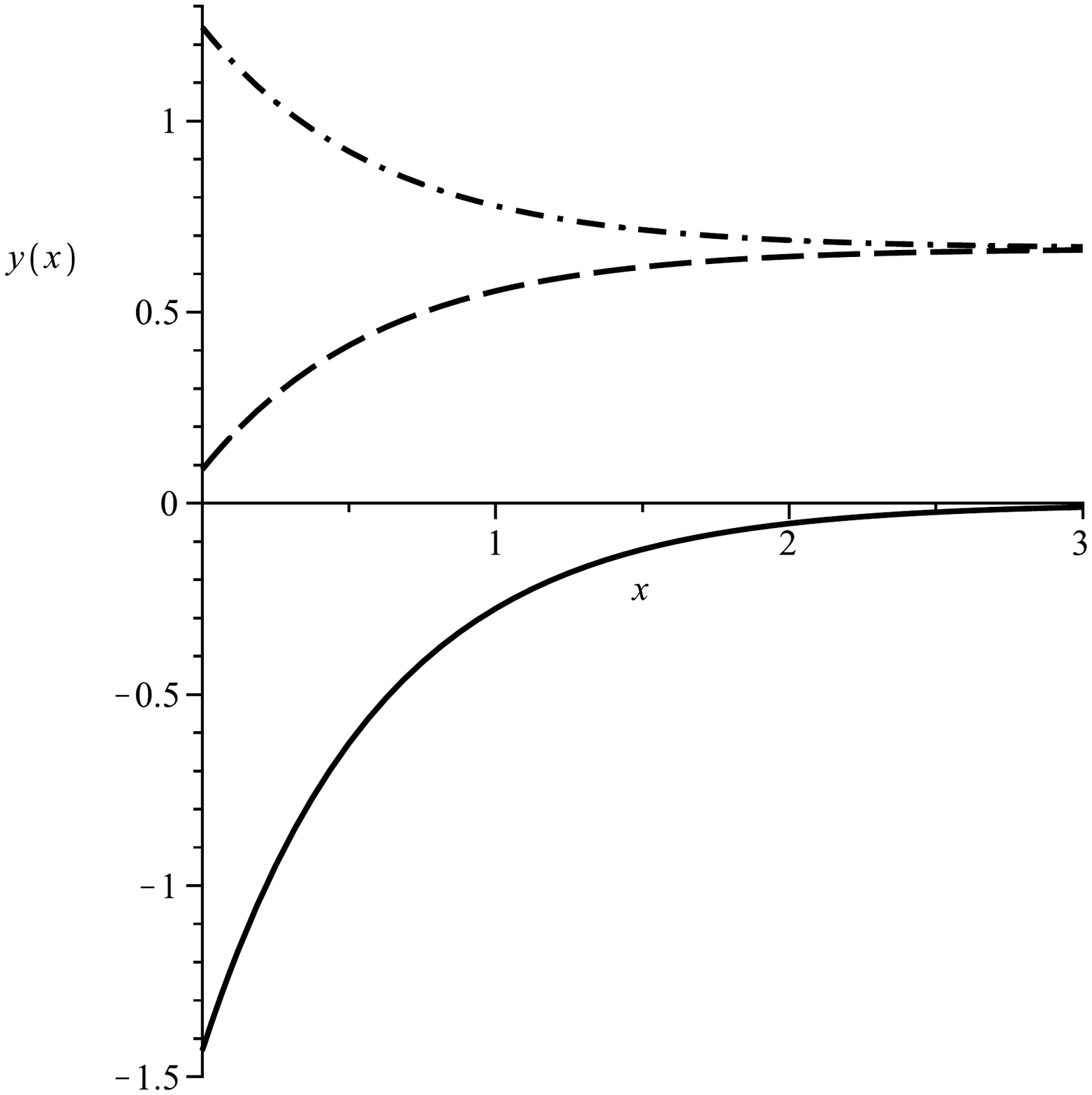}
}}
\caption{Graphs of $c_+(x)$ (dash), $c_-(x)$ (dashdot), $E(x)$ (solid), $A_+$ (dot) and $A_-$ (spacedot), in the left reservoir, 
in the slab, and in the right reservoir, 
corresponding to corrected (approximate) BCs.  Parameter values are  $c_{(-\infty)}=1/3$, $c_{(+\infty)}=2/3$, $\lambda=0.7$, $\alpha_+=0.4$,
$\alpha_-=0.6$, 
with $j_0=0.16$.}
\end{figure}

This boundary value problem can also be posed in terms of PII, which in  this case takes a simpler, more standard form than 
\eqref{painleve_neumann}.   To see this, note firstly that  $c_0$ and $c_1$ in \eqref{painleve_neumann} 
are now to be replaced by 
$c_{(-\infty)}$ and $c_{(+\infty)}$, respectively, according to the first of \eqref{reservoir10} and the first of \eqref{reservoir12}. 
Then note that, as a consequence of \eqref{reservoir5} and the third of \eqref{reservoir8},
\bea
\half\lambda^2\, E(0)^2\ll 2c_{(-\infty)}\,.
\label{reservoir14}
\eea
Similarly
\bea
\half\lambda_1^2 E(1)^2\ll c_{(+\infty)}\,.
\label{reservoir15}
\eea
The terms on the left-hand sides of these inequalities are therefore to be neglected relative to the terms on 
the right-hand sides in the ODE that replaces \eqref{painleve_neumann}, and it therefore takes the form
\bea
\lambda^2\,E\,''(x)=\half\lambda^2E(x)^3
+2\left[c_{(-\infty)}+\left\{c_{(+\infty)}-c_{(-\infty)}\right\}x\right]E(x)
\mea\mea
-2(\alpha_+-\alpha_-)\left\{c_{(+\infty)}-c_{(-\infty)}\right\}-2j_{\,0}\,.
\label{painleve_radiation}
\eea
This equation is to be solved on $0<x<1$, subject to the radiation BCs \eqref{reservoir11} and \eqref{reservoir13}. The constants
 $c_{(-\infty)}$, $c_{(+\infty)}$, $\alpha_+$, $\alpha_-$ and $j_{\,0}$ appearing here are to be regarded as given. Nothing is known about 
 existence or uniqueness of solutions to this problem, to our knowledge, nor about positivity of resultant concentrations.  

Given \eqref{reservoir14} and \eqref{reservoir15}, the question arises if $\half\lambda^2 E(x)^2$ should also be neglected relative to 
$c_{(-\infty)}$ and $c_{(+\infty)}$ for all $0<x<1$, and not only at the boundaries.  If so, then the cubic term in \eqref{painleve_radiation} drops out, and the resultant linear boundary-value problem  should be  exactly solvable in terms of Airy functions ({\em cf.} \cite{bass2}).   
This will not be pursued here.

Similar manipulations to those in the preceding subsection 
show that, like charge-neutral BCs, those in \eqref{reservoir10}, \eqref{reservoir12} and \eqref{currentBC} are also not preserved 
in form under B\"acklund transformations and their inverses.
\section{Concluding remarks}
Quantization of ionic fluxes and of the electric current-density by B\"acklund transformations of solutions, 
in 
this one-dimensional model of electrodiffusion within a liquid junction, is a remarkable mathematical phenomenon.  
What physical significance can be attached to this phenomenon is not completely clear, but the example provided by the exact 
Planck solution and its transforms
is highly suggestive of a corresponding ground state and excited states of the system.  A difficulty with this interpretation, and with
the concept of flux-quantization by B\"acklund transformations more generally, is that BCs that are invariant in form 
under such transformations have not yet been identified.  

We have seen that  familiar charge-neutral BCs, first introduced in the context of electro-diffusion by Planck, 
lead to  several attractive properties
of solutions.  Existence of such solutions has previously been proved under certain conditions \cite{thompson,amster}, and we have shown 
positivity of the functions
representing ionic concentrations.  However, we have argued that such BCs are not physically consistent.  Moreover, they are not preserved in 
form under B\"acklund transformations.  This has prompted us to examine more realistic BCs.  To find these we have had to solve for the exact form 
of ionic concentrations and the electric field in well-stirred reservoirs outside each face of the junction.   The new BCs define novel boundary-value
problems for the system of ODEs
that determine the model of electrodiffusion in the junction.  A physically sensible approximation to these 
reservoir solutions has been found to lead to radiation BCs for   the Painlev\'e II ODE which 
lies at the heart of the model, while the exact solutions lead to highly nonlinear BCs for that ODE.

These new BCs, either exact or approximate, are not preserved in form under B\"acklund transformations, and it  remains to classify BCs, as 
generated by such transformations, that are physically relevant. 
The application of B\"acklund transformations to link boundary value problems of physical relevance does arise elsewhere,
in particular in elasticity \cite{rogers7}.  There mixed boundary value problems associated with normal loading and application of 
torsion may be  linked in that way.  Further applications involve boundary value problems describing indentation of 
shear-strained nonlinear elastic materials \cite{rogers8}.  We can be optimistic, therefore, that B\"acklund transformations can also 
be shown to relate physically interesting BCs in the present context.  

We have shown for the present model
that B\"acklund transformations do leave invariant two quantities.  
One represents the total ionic (osmotic) pressure within the junction. The other represents the total 
resistive force per unit volume that the solute offers against ionic transport. We may hope that the invariance of these 
quantities provides a clue in the ongoing study of BCs and their relationship to flux-quantization by B\"acklund transformations.      
\renewcommand{\theequation}{A\arabic{equation}}
\setcounter{equation}{0}
\section*{Appendix A:  Gambier transformations of exact solutions}
Consider again the exact case of Planck's solution, as in \eqref{ground}.  Note that for this solution,
$B$ as in \eqref{first_integral} takes the value $2c_0$, and $A_+=A_-=A$, as required for application of ${\cal G}^{-1}_+$.  
We consider 
a small perturbation of this solution in the form
\bea
c_{\pm}(x)= c_0 +Ax+ \epsilon\, d_{\pm}(x)\,,\quad E(x)=\epsilon\, F(x)\,,\quad A_{\pm}=A\,,
\label{airypert1}
\eea 
where $d_{\pm}(x)$ and $F(x)$ are to be determined by requiring that $c_{\pm}(x)$, $E(x)$ and $A_{\pm}$ satisfy
\eqref{system2} to first-order  in $\epsilon$, given that they satisfy the system to zeroth-order because \eqref{ground} 
is already a solution.  Substituting from \eqref{airypert1} into \eqref{system2} gives at first-order 
\bea
d_+\,'(x)=(c_0 +Ax)\,F(x)\,,\quad  d_-\,'(x)=-(c_0 +Ax)\,F(x)\,,
\mea\mea
\lambda^2 F\,'(x)=d_+(x)-d_-(x)\,,\qquad\qquad\qquad
\label{airypert2}
\eea
and hence
\bea
\lambda^2 F\,''(x)=2(c_0 +Ax)\,F(x)\,.
\label{airypert3}
\eea
This has the general solution
\bea
F(x)=a {\rm Ai}(s)+ b {\rm Bi}(s)\,,\quad s=2(c_0 +Ax)/(4\lambda^2 A^2)^{1/3}
\label{airypert4}
\eea
where Ai and Bi are Airy functions of the first and second kind \cite{abram}, and $a$, $b$ are arbitrary constants.  
Substituting from \eqref{airypert1} into \eqref{gambier3} after noting that 
\bea
c_+(x)-c_-(x)&=&\epsilon [d_+(x)-d_-(x)]
\mea\mea
&=&\epsilon\lambda^2  F\,'(x)
\mea\mea
&=&\epsilon (J\lambda^4)^{1/3}[a {\rm Ai}'(s)+b {\rm Bi}'(s)\,,
\label{airypert5}
\eea
and then letting $\epsilon\to 0$, we obtain the new solution
\bea
{\hat c}_+(x)=0\,,\quad
{\hat c}_-(x)=2\lambda^2 F'(x)^2/F(x)^2-4Ax-4c_0
\mea\mea
{\hat E}(x)= 2F'(x)/F(x)\,,\quad {\hat A}_+=0\,,\quad {\hat A}_-=-4A\,,\quad {\hat B}=-4c_0\,,
\label{airypert6}
\eea
with $F$ and $F'$ as in \eqref{airypert4} and \eqref{airypert5}.  The treatment of ${\cal G}_-^{-1}$, 
from the same starting point, is similar, and leads to the conjugate solution to \eqref{airypert6}, as follows from \eqref{conjugate_gambier}.  
It is easily checked also that, just as the inverse Gambier transformations carry the seed solution \eqref{ground} for the sequence 
of exact rational solutions of PII into the seed solution for the sequence of exact Airy solutions, so the Gambier transformations 
${\cal G}_{\pm}$ carry the seed solutions  for these Airy sequences, 
namely \eqref{airypert6} and its conjugate, back into the seed solution \eqref{ground} for the sequence of rational solutions.

{\em Remark}: These calculations reveal, for the first time, a connection between the approximate solution \cite{bass2} of PII in terms of Airy functions, and the known sequences of exact solutions of PII in terms of Airy functions \cite{rogers6}.  The former is essentially
$E(x)$ as in \eqref{airypert1}, \eqref{airypert4}, while the latter are  generated from \eqref{airypert6} and its conjugate by B\"acklund transformations. 
\renewcommand{\theequation}{B\arabic{equation}}
\setcounter{equation}{0}
\section*{Appendix B:  Exact boundary conditions and reservoir solutions}
The exact solution of \eqref{reservoir4}, with the behaviour at $x=-\infty$ described there, is known to be \cite{herzfeld,langmuir}
\bea
\varphi(x)= 2\ln\left[\left(1+Ae^{x/\lambda_0}\right)/\left(1-Ae^{x/\lambda_0}\right)\right]\,,\quad -1<A<1\,,
\label{B10}
\eea
where $\lambda_0$ is as in \eqref{reservoir6}.  
From \eqref{reservoir2} and \eqref{reservoir3} we then get
\bea
c_+(x)=c_{(-\infty)}\left[\left(1-Ae^{x/\lambda_0}\right)/\left(1+Ae^{x/\lambda_0}\right)\right]^2\,,
\mea\mea
c_-(x)=c_{(-\infty)}\left[\left(1+Ae^{x/\lambda_0}\right)/\left(1-Ae^{x/\lambda_0}\right)\right]^2\,,
\mea\mea
E(x)=-4Ae^{x/\lambda_0}/\lambda_0\left(1-A^2e^{2x/\lambda_0}\right)\,.
\label{B11}
\eea
throughout the left-hand reservoir.  The approximations are recovered when $\exp(x/\lambda_0)\ll 1$, so that \eqref{reservoir8} hold as before.  

From \eqref{B11} we have in particular that 
\bea
c_+(0)=c_{(-\infty)}\,(1-A)^2/(1+A)^2\,,
\mea\mea
c_-(0)=c_{(-\infty)}\,(1+A)^2/(1-A)^2\,,
\mea\mea
E(0)=-4A/\lambda_0(1-A^2)\,,
\label{B12}
\eea
from which we get
\bea
c_+(0)c_-(0)=c_{(-\infty)}\,^2\,,\quad \lambda_0 E(0)=\sqrt{2c_+(0)}-\sqrt{2c_-(0)}\,.
\label{B13}
\eea
These are the nonlinear 
exact BCs that now apply at the left-hand reservoir-junction interface when solving \eqref{system2}.  Similarly, at the right-hand interface
we get 
\bea
c_+(1)c_-(1)=c_{(+\infty)}\,^2\,,\quad \lambda_1 E(1)=\sqrt{2c_+(1)}-\sqrt{2c_-(1)}\,.
\label{B14}
\eea 
With a little effort  it can be seen that the conditions \eqref{B13} and \eqref{B14}   can also be written in the form
\bea
c_{\pm}(0)=c_{(-\infty)}+\quarter\lambda^2  E(0)^2\pm \quarter \lambda E(0)\sqrt{8c_{(-\infty)}+\lambda^2 E(0)^2}\,,
\mea\mea
c_{\pm}(1)=c_{(+\infty)}+\quarter\lambda^2  E(1)^2\pm \quarter\lambda  E(1)\sqrt{8c_{(+\infty)}+\lambda^2 E(1)^2}\,,
\label{B15}
\eea
from which the approximate forms \eqref{reservoir8} are easily recovered.

The nonlinear BCs  that apply for PII in the slab in this case can be seen from \eqref{B15}; 
nothing is known about existence or uniqueness of its solutions. 

Once again it can be checked that
BCs of the form \eqref{B15} are  not preserved during flux-quantization by  B\"acklund transformations.

\end{document}